\documentclass[twocolumn]{aastex63}
\usepackage{graphicx}
\usepackage{amsmath}
\usepackage{subfigure}
\usepackage{multirow}
\usepackage{soul}

\begin{document}

\title{ZTF IC 10 variable catalog}
\author{Zehao Jin}
\author{Joseph D. Gelfand}
\affiliation{New York University Abu Dhabi, PO Box 129188, Abu Dhabi, UAE}
\affiliation{Center for Astrophysics and Space Science  (CASS), NYU Abu Dhabi, PO Box 129188, Abu Dhabi, UAE}

\date{\today}

\begin{abstract}

To study how massive variable stars effect their environment, we search for variability among Zwicky Transient Facility (ZTF) sources located within the optical extent of a nearby starburst galaxy IC 10. 
We present the ZTF IC 10 catalog, which classifies 1516 $r$ band sources and 864 $g$ band sources within a $225''$ radius around IC 10 into three categories: 1388 (767) $r$ ($g$) band non-variables, 150 (85) $r$ ($g$) band non-periodic variables, and 37 (12) $r$ ($g$) band periodic variables. 
Among them 101 (48) $r$ ($g$) band non-periodic variables, and 22 (4) $r$ ($g$) band periodic variables are inside IC 10. 
We verify our classification by cross-matching with previous variability catalogs and machine learning powered classifications. 
Various analysis including population demographics, color-magnitude diagrams, and cross matching with a set of different surveys and database such as Gaia, XMM-Newton, Chandra, and SIMBAD are also presented. 
Based on source density and parallax, we distinguish sources within IC 10 from non-IC 10 sources. 
For IC 10 sources, we highlight flaring super giants, a source with long secondary period, periodic super giants including possible S Doradus luminous blue variable and candidate Miras. 
For non-IC 10 sources, we present super red sources and compact objects such as a possible long period subdwarf and a periodic X-ray source. 
The catalog can serve as a useful database to study the connection between various type of massive stars and their host galaxies.
\end{abstract}

\section{Introduction}
Understanding galaxy evolution is a crucial part of modern astrophysics. One contributor to galaxy evolution is massive stars, which power stellar feedback \citep{Ceverino_2009}. Massive stars drive stellar feedback in very different ways, from violent supernova explosions, to stellar winds, transients, and ejections \citep{Hopkins_2011,Hopkins_2012,Schneider_2020}. The optical emission of massive stars are expected to be variable during their late evolutionary stages, and this variability is usually related to the physical processes responsible for stellar feedback \citep{Richardson_2011,Kourniotis_2014,Anders_2023}. The exact mechanism of stellar feedback is strongly related to the type of massive stars, therefore a rich and diverse population of massive stars is needed to study how they collectively affect their environment.

Nearby starburst galaxy IC 10 provides an unique setting studying massive stars. It contains a large number of different types of massive star systems, such as Luminous Blue Variables, Wolf-Rayet stars, and High-Mass X-ray binaries \citep{Bauer_2004,2005MNRAS.362.1065W}. Furthermore, IC 10 is a dwarf galaxy with low metallicity, similar to galaxies in the early universe.

As the first step to better study massive stars in IC 10, this work uses Zwicky Transient Facility (ZTF) $r$ band and $g$ band data to identify optical transients in IC 10. We provide a catalog with the variability information useful for determining the nature of the source, as outlined in \cite{Smith_2011}, as well as a series of analysis on the variability statistics to highlight interesting sources.

Early work on variable stars in IC 10 began with the pioneering study by \cite{Saha_1996}, who analyzed repeated images collected over an 11-year span to identify candidate variables in this starburst dwarf galaxy, including four Cepheids, from which a distance modulus has been derived. Recently, \cite{Gholami_2021} used the Isaac Newton Telescope (INT) to employed optical monitor to identify 762 variable candidates including 424 long-period variables. These variables were used to reconstruct the galaxy's star formation history, confirming IC 10 is currently undergoing high levels of star formation. Complementary to these optical efforts, Chandra X-ray observations by \cite{Laycock_2017} uncovered 21 variable X-ray sources, potentially High Mass X-ray Binaries (HMXBs), in IC 10, including four of them with a strong variability. While not IC 10-specific, automated transient search facilities such as ZTF also revealed variable stars in IC 10. In a ZTF variable catalog by \citep{Chen_2020} based on ZTF DR2, 781,602 periodic variable stars are identified, and one of them falls in the field of IC 10. Shortly after, \cite{Cheung_2021} proposed a new model to further classify these periodic variables in Chen's catalog using a convolutional variational autoencoder and hierarchical random forest. This work is dedicated to present a catalog of IC 10 non-periodic variable or periodic variable stars in ZTF $r$ and $g$ band. By focusing of purity of the sample, carefully handling foreground and background sources, and cross-checking with many other surveys, we provide a list of objects that worth follow-up investigations.

In \S\ref{sec:data} we introduce the data used in this work. \S\ref{sec:method} details and verifies our classification scheme. In \S\ref{sec:demo} we present the statistics of our catalog in various aspects, including using Gaia to distinguish potential IC 10 sources and foreground sources. In \S\ref{sec:ic10sources} we focus on sources within IC 10, while in \S\ref{sec:othersources} we explore foreground sources, and we conclude in \S\ref{sec:conclusion}. 

The ZTF IC 10 variable catalog and related scripts are publicly are available on GitHub\footnote{GitHub repository: \url{https://github.com/ZehaoJin/Transients-in-IC-10}} and the catalog is also archived in Zenodo\citep{jin_2025_zenodo}.

\section{Data}
\label{sec:data}

\begin{table*}
\centering
\begin{tabular}{ll}
Filter info   & ZTF $r$ band: $\lambda_{cen}=$ 6434 {\AA}, $\mathrm{FWHM}=1557$ {\AA}, Limiting magnitude 20.6 mag     \\
              & ZTF $g$ band: $\lambda_{cen}=$ 4803 {\AA}, $\mathrm{FWHM}=1321$ {\AA}, Limiting magnitude 20.8 mag     \\ \hline
IC 10 field   & $225''$ radius circle around $\mathrm{RA}=00:20:23.16,\ \mathrm{DEC}=+59:17:34.7$ \\ \hline
Dates covered & ZTF DR15, 2018 Mar - 2022 Nov (MJD 58194 - 59892)         
\end{tabular}
\caption{\label{tab:data}Summary of data used in this paper}
\end{table*}

A brief summary of data used in this work is presented in Table \ref{tab:data}.

The Zwicky Transient Facility (ZTF) is a wide-field optical time-domain survey that began operations in March 2018 \citep{Bellm_2019}. ZTF aims to discover and study transient and variable astrophysical sources, such as supernovae, active galactic nuclei, and asteroids, among others\footnote{ZTF data \citep{2018} are publicly available through the Infrared Science Archive (IRSA) \url{https://irsa.ipac.caltech.edu/Missions/ztf.html}.}. ZTF has observed IC 10 roughly everyday during the past few years, making it well suited for identifying variables. 
Here we use the ZTF DR 15 data in $g$ band and $r$ band, since they are two most sensitive ZTF bands. 
We do a radial query of $225^{\prime \prime}$ around the center of IC 10 ($\mathrm{RA}=00:20:23.16,\ \mathrm{DEC}=+59:17:34.7$ \citep{1999ApJS..125..409C}) to cover the optical size of IC 10 \citep{Huchra_1999}. 2407 and 1334 ZTF labeled lightcurves are found in $r$ and $g$ band, respectively. The lightcurves cover roughly a four-year span starting from March 2018. Only clear observations without any flags are used in this work.

\section{Methodology}
\label{sec:method}
\begin{figure}
\centering
\includegraphics[width=0.9\linewidth]{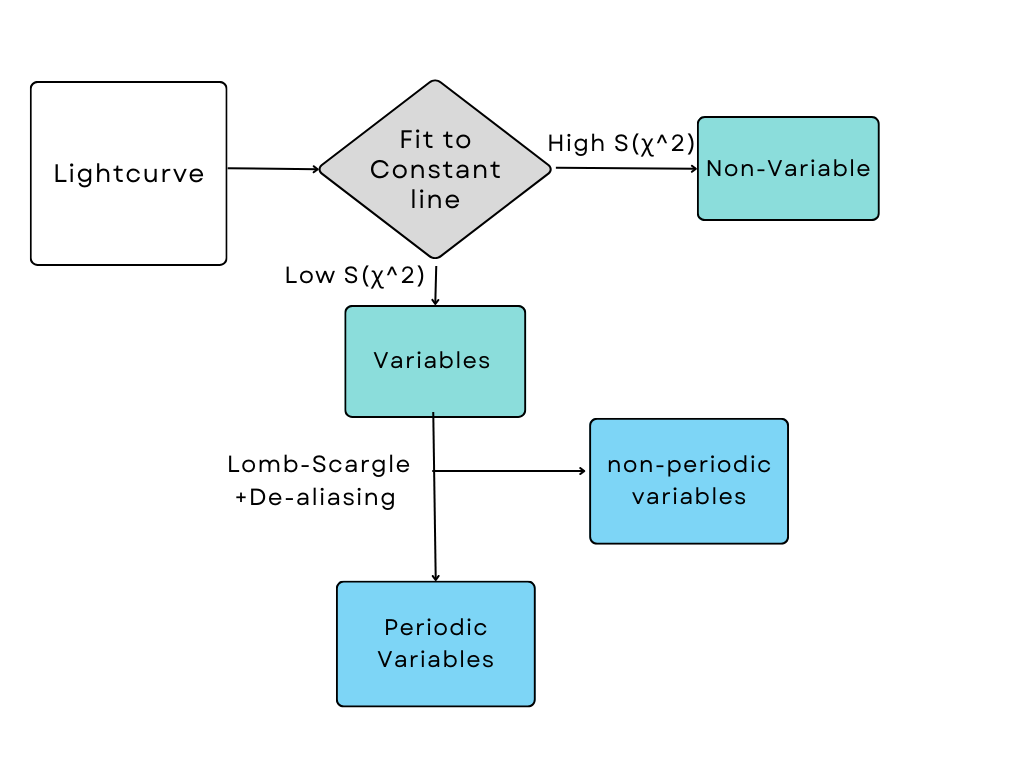}
\caption{Visualization of our methodology}
\label{fig:flow}
\end{figure}

For every lightcurve, we first determine if it is variable by fitting it to a constant magnitude and calculating the value of the survival function $S_k(\chi^2)$ of this model (\S\ref{sec:variabilitycut}). For variables, we apply Lomb-Scargle analysis and determine whether it is periodic or non-periodic by using the false alarm probability (FAP, see \S\ref{sec:FAP}), and checks for aliasing with $Q_P$ (\S\ref{sec:Q_P}) and $C_T$ (\S\ref{sec:C_T}). In the end, we group together OIDs (ZTF source label) whose positions are $<1''$ apart, since these are likely the same astronomical source (\S\ref{sec:group}). A summary of our criteria can be find in flowchart (Figure \ref{fig:flow}) and Table \ref{tab:cut}, and these steps are described in detail below.

\begin{table*}
\centering
\begin{tabular}{llll}
      & Definition & Cutoff in $r$ band & Cutoff in $g$ band \\ \hline
$S_k(\chi^2)$ & Survival function of the constant magnitude model & $S_k(\chi^2)<4.15 \times 10^{-6}$ & $S_k(\chi^2)<7.50 \times 10^{-6}$ \\
  & (Variables/All lightcurves in band) & (250/2407) & (117/1334) \\ \hline
$\rm FAP$ & False alarm probability from Lomb-Scargle & $\rm FAP<4 \times 10^{-5}$ & $\rm FAP<8.55 \times 10^{-5}$ \\
  & (Low FAP/Variables) & (106/250) & (44/117) \\ \hline
$Q_{P}$  & $P_{\rm LS}/P_{\rm window}$ & $Q_{P}>1$ & $Q_{P}>1$ \\
  & (Low FAP, High $Q_{P}$/Low FAP) & (66/106) & (27/44) \\ \hline
$C_{T}$ & $|T_{\rm LS}-T_{\rm window}|/T_{\rm window}$ & $C_{T}>0.01$ & $C_{T}>0.01$ \\
  &  & $C_{T}<0.49, \ C_{T}>0.91$ & $C_{T}<0.49, \ C_{T}>0.91$ \\
  & (Periodic variables/Low FAP, High $Q_{P}$) & (38/66) & (12/27) \\ \hline
  & Merge nearby lightcurves into the same source & $d<1''$ & $d<1''$ \\
  & (Periodic variable sources/Periodic variable lightcurves) & (37/38) & (12/12) \\
  & (Non-periodic variable sources/Non-periodic variable lightcurves) & (150/212) & (85/105) \\
  & (Non-variable sources/Non-variable lightcurves) & (1329/2157) & (767/1217) \\
  & (All sources in band/All lightcurves in band) & (1516/2407) & (864/1334) 
\end{tabular}
\caption{\label{tab:cut}Periodic variable criteria. The fractions in parentheses denotes the (subcategory/parent category) the cut acts on, and the number (before/after) the cut.}
\end{table*}

\subsection{Variability cut \texorpdfstring{$S_k(\chi^2)$}{S\_k(chi\^2)}}
\label{sec:variabilitycut}
A non-variable source by definition has a constant magnitude with time. By evaluating the fit of the lightcurve to a constant value, we can effectively separate possible variable sources from non-variable sources. For a lightcurve with $N$ observations of magnitudes $m_i$ and errors in magnitude $\sigma_i$ , the chi-square $\chi^2$ of the best fitting constant magnitude $m_C$ is

\begin{equation}
\chi^2=\sum_{i=1}^n\frac{(m_i-m_C)^2}{\sigma_i^2}
\end{equation}

\begin{figure*}
\centering
\subfigure{
\includegraphics[width=0.45\linewidth]{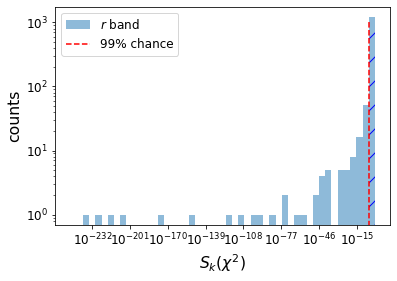}}
\subfigure{
\includegraphics[width=0.45\linewidth]{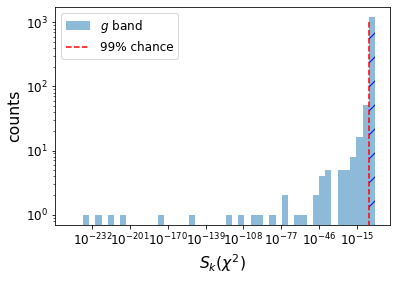}}
\caption{Distribution of $S_k(\chi^2)$ for all lightcurves in ZTF $r$ and $g$ band. Lightcurves left to the red dotted line has 99\% or more chance to be variable and is classified as variable. Hatched region (right to the red dotted line) is excluded.}
\label{fig:sf}
\end{figure*}

For a non-variable source, changes in $m$ will be dominated by statistical (measurement) error $\sigma_i$, but not intrinsic variations, therefore a constant magnitude $m_C$ fit should give $\chi^2\approx N$. The cumulative distribution function (CDF) is the probability that the $\chi^2$ would by chance exceed a particular value for $k=N-1$ degrees of freedom. We hence measure the probability that an observed lightcurve matches the constant line model via survival function $S_k(\chi^2)$, the chance that a constant line fits the observed lightcurve,
\begin{equation}
S_k(\chi^2)=1-\mathrm{CDF}_k(\chi^2)
\end{equation}
where $\mathrm{CDF}_k(\chi^2)$ is the cumulative density function of a $\chi^2$ distribution with degree of freedom $k$. To make sure the chance of even one ``variable'' lightcurve being mislabeled as variable is below 1\%, we identify variable lightcurves as those with
\begin{equation}
S_k(\chi^2)\leq \frac{1}{100 \times N_{lc}}
\end{equation}
where $N_{lc}$ is the total number of lightcurves in each band. The rest are classified as non-variable lightcurves. The distribution of $S_k(\chi^2)$ and its cutoff can be visualized in Figure \ref{fig:sf}. As shown in Table \ref{tab:cut}, we identify 250 out of 2407 lightcurves in $r$ band, and 117 out of 1334 lightcurves in $g$ band as variable based on their low values of $S_k(\chi^2)$. 

\subsection{Classify variables with periodicity}
\label{sec:periodicity}
Once we identify variable lightcurves (\S\ref{sec:variabilitycut}), we apply Lomb-Scargle periodogram based criteria FAP, $Q_P$, and $C_T$ to further determine if the variability is periodic or non-periodic. The folded lightcurves of all periodic lightcurves are publicly available at \url{https://github.com/ZehaoJin/Transients-in-IC-10/blob/main/All_periodic_lightcurves.ipynb}.

\subsubsection{\texorpdfstring{FAP}{FAP}}
\label{sec:FAP}
The Lomb-Scargle periodogram is a commonly used tool to characterize periodic signals in unevenly spaced observations developed by \cite{1976Ap&SS..39..447L} and \cite{1982ApJ...263..835S}. A periodogram calculates the power on a frequency/period grid, while the false alarm probability ($\rm FAP$), gives the probability of measuring a peak at a certain or higher power assuming Gaussian noise and non-periodic data. A smaller FAP suggests that emission from the source is more likely to be periodic. 

\begin{figure*}
\centering
\subfigure{
\includegraphics[width=0.45\linewidth]{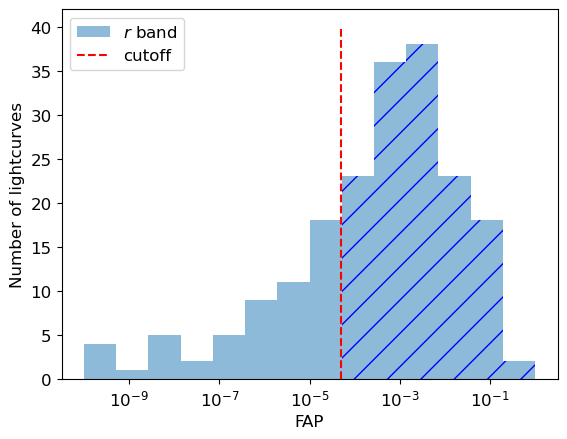}}
\subfigure{
\includegraphics[width=0.45\linewidth]{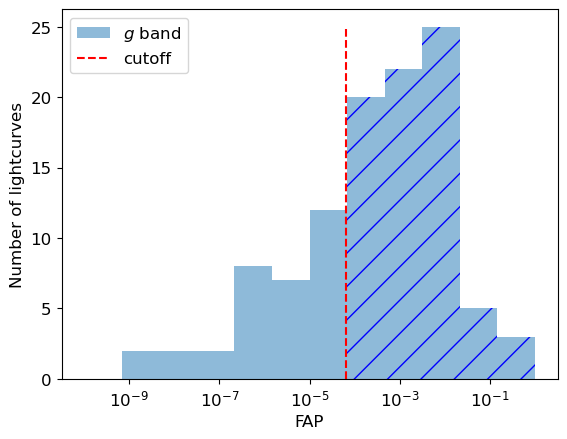}}
\caption{Distribution of FAP for variables in ZTF $r$ and $g$ band. We keep lightcurves left to the red dotted line as low FAP lightcurves. Hatched region (right to the red dotted line) is excluded.}
\label{fig:fap}
\end{figure*}

Here we use Python package \emph{Astropy} \citep{astropy:2013,astropy:2018} to calculate the Lomb-Scargle periodogram, which approximates the FAP of the Lomb-Scargle periodogram peak following the method developed by \cite{ 2008MNRAS.385.1279B}. The distribution of FAP for variables in $r$ and $g$ band is shown in Figure \ref{fig:fap}. To produce a list of periodic lightcurves that has a probability $\leq1\%$ of one misclassified periodic lightcurves, we set the FAP cutoff to $\frac{1}{100 \times N_{lc,var}}$, where $N_{lc,var}$ is the total number of variable lightcurves in the corresponding band. Specifically, we keep lightcurves with FAP lower than the cutoff (Table \ref{tab:cut}) as ``low FAP" lightcurves. 

Although FAP is one of the most effective criteria to detect periodic signals, it is vulnerable to aliasing. The nightly/quarter/yearly observation window pattern can create fake peaks and contaminate the real peak on the periodogram. The following two criteria are mainly devoted to the de-aliasing of low FAP lightcurves.

\subsubsection{\texorpdfstring{$Q_{P}$}{Q\_P}}
\label{sec:Q_P}
\begin{figure*}
\centering
\subfigure{
\includegraphics[width=0.45\linewidth]{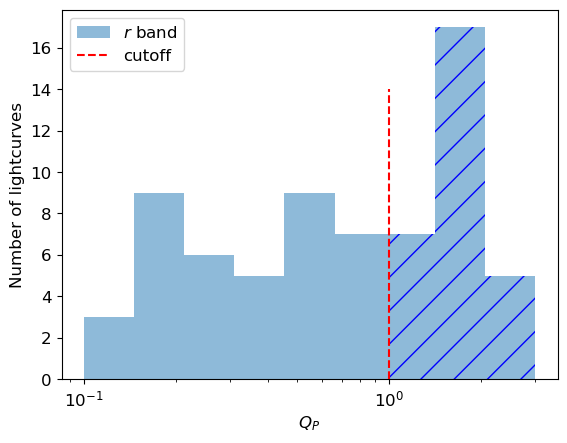}}
\subfigure{
\includegraphics[width=0.45\linewidth]{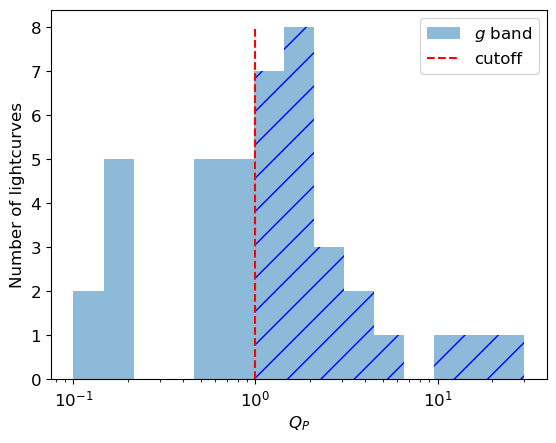}}
\caption{Distribution of $Q_{P}$ for low $\rm FAP$ lightcurves in ZTF $r$ and $g$ band. We keep lightcurves right to the red dotted line, remove the ones on the left. Hatched region (right to the red dotted line) is excluded.}
\label{fig:qp}
\end{figure*}

Periodicity can sometimes appear due to spacing of observations, but not intrinsic changes in the luminosity of a source. To avoid aliasing, a common approach is to look at the window function, where one calculates the Lomb-Scargle periodogram of a lightcurve with constant magnitude and same observation dates. We then define $Q_{P}$ as the ratio of powers, 

\begin{equation}
Q_{P}=P_{\rm LS}/P_{\rm window}
\end{equation}

where $P_{\rm LS}$ is the Lomb-Scargle periodogram peak power, and $P_{\rm window}$ is the window function power at the period of the original Lomb-Scargle periodogram peak. $Q_{P}>1$ indicates the period or frequency we found is more significant in the lightcurve of interest than for a constant magnitude model given the same spacing in observations. An intrinsically non-periodic source could have a low FAP due to aliasing, but it not likely to have a $Q_{P}>1$. Therefore such a cut will help removing aliasing. The distribution of $Q_{P}$ for low FAP lightcurves in both bands is shown in Figure \ref{fig:qp}.

\subsubsection{\texorpdfstring{$C_{T}$}{C\_T}}
\label{sec:C_T}
\begin{figure*}
\centering
\subfigure{
\includegraphics[width=0.45\linewidth]{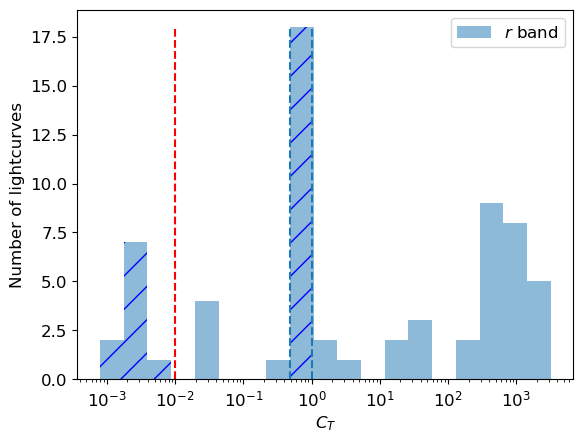}}
\subfigure{
\includegraphics[width=0.45\linewidth]{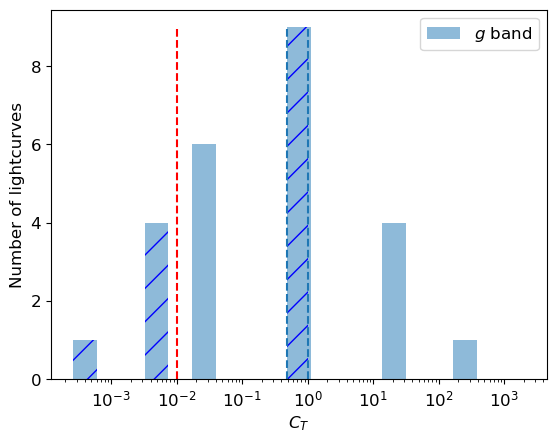}}
\caption{Distribution of $C_{T}$ for low FAP, high $Q_P$ lightcurves in ZTF $r$ and $g$ band. We cut (hatched region) lightcurves left to the red dotted line, and lightcurves in between blue dotted lines, with exception of extremely low $\rm FAP$.}
\label{fig:ct}
\end{figure*}

$C_{T}$ is the relative change in periods, 
\begin{equation}
C_{T}=|T_{\rm LS}-T_{\rm window}|/T_{\rm window}
\end{equation}
where $T_{\rm LS}$ is the period at Lomb-Scargle periodogram peak power, $T_{\rm window}$ is the period at window function peak power. $C_{T}$ tells how close the period we found is to the aliasing period. We thus rule out lightcurves with $C_{T}<0.01$, i.e. period and aliasing period only differ by $1\%$. As Figure \ref{fig:ct} shows, there is a spike between $C_{T}=$ 0.5 and 1 in both bands. Due to the fact that ZTF make observations on a daily basis, most of lightcurves have window function period $T_{\rm window}=1 \ {\rm day}$. Aliases appear following 
\begin{equation}
\frac{1}{T_{\rm alias}}=\frac{1}{T_{\rm true}}+\frac{n}{T_{\rm window}}
\end{equation}
for integer values of $n$. In cases with no true period ($T_{\rm true}\rightarrow\infty$ and $T_{\rm LS}=T_{\rm alias}$), we get
\begin{equation}
\frac{1}{T_{\rm alias}}=\frac{n}{T_{\rm window}},
\end{equation}
or 
\begin{equation}
\frac{T_{\rm alias}}{T_{\rm window}}=\frac{1}{n},
\end{equation}
then
\begin{equation}
C_{T}=|\frac{T_{\rm alias}}{T_{\rm window}}-1|=|\frac{1}{n}-1|.
\end{equation}
For $n=2 \rightarrow \infty$, $C_{T}=1/2 \rightarrow 1$, exactly where the spike is located. To resolve such aliasing, we rule out lightcurves with $0.5 \leq C_{T} \leq 1$. To prevent removing real periodic lightcurves with periods shorter than 0.5 days, we preserve lightcurves that has extremely low FAP (FAP $<10^{-30}$, happened only once).

\subsection{Group nearby OIDs}
\label{sec:group}

According to \cite{Bellm_2019}, ZTF has a median seeing (FWHM) $\sim 2''$. While this is greater than the $<0.1''$ astrometric positional uncertainty of this survey \citep{Masci_2019}, emission from sources $<1''$ will be blended together, so light curves of sources so close together will be likely a combination of both. 
Therefore we group OIDs less than $1''$ apart as one astronomical source, and find that in most (59\% and 54\% of astronomical sources in $r$ and $g$ band) cases, each source has two OIDs associated with it, one with $<100$ observations, the other with a much higher number. In our catalog the variability and periodicity of a sources with multiple OIDs is determined by the OID with highest number of observations. If a source's highest number of observation OID do not shown any variability, but has an OID with lower number of observations that is variable, a variable suspect flag will be raised in the catalog, although such cases are very rare (1 in $r$ band and 4 in $g$ band). We do not place a periodic suspect flag since any periodicity found in OID with smaller number of observations but not in OID with higher number of observations is unlikely to be true. Note that after the $1''$ grouping, most of the grouped OIDs should be the same source, but there is the chance they are different sources very close to each other on the sky. To account for this issue, we match our catalog with Gaia, which has a higher angular resolution that can resolve sources within $1''$. A more detailed discussion of Gaia counterparts can be found in \S \ref{sec:gaia}.

\begin{figure*}
\centering
\subfigure{
\includegraphics[width=0.45\linewidth]{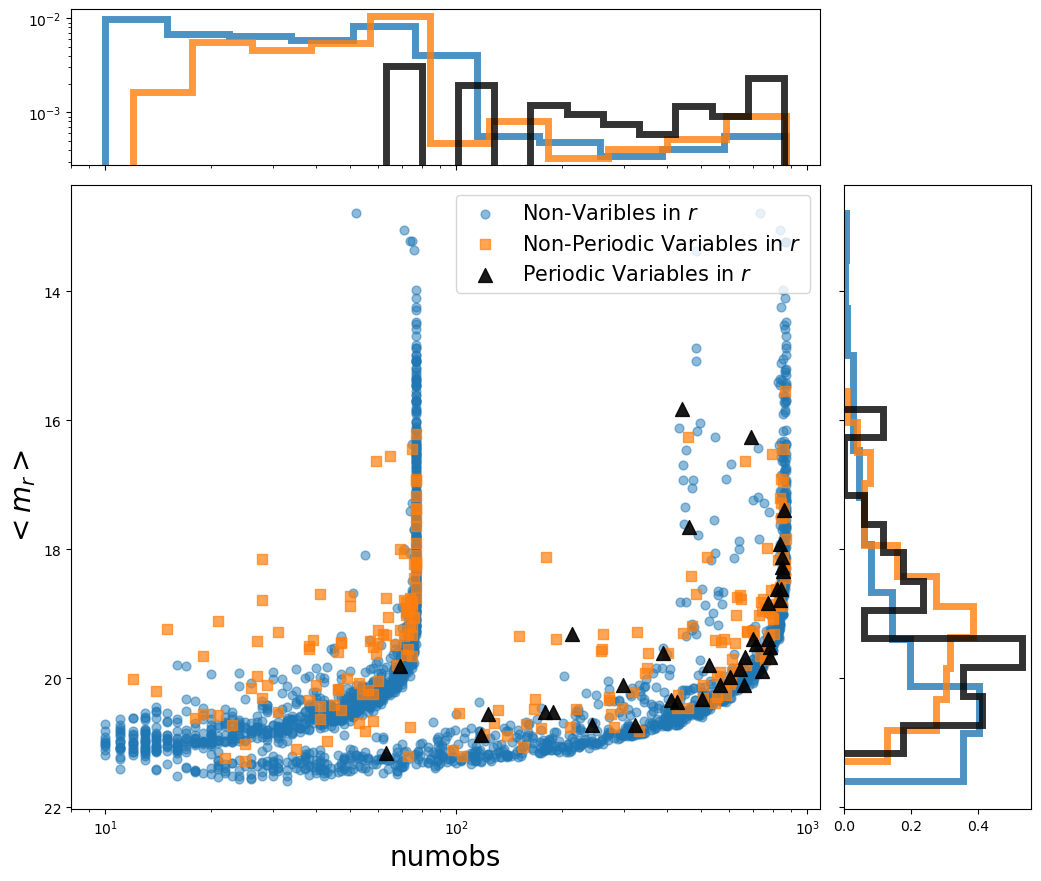}}
\subfigure{
\includegraphics[width=0.45\linewidth]{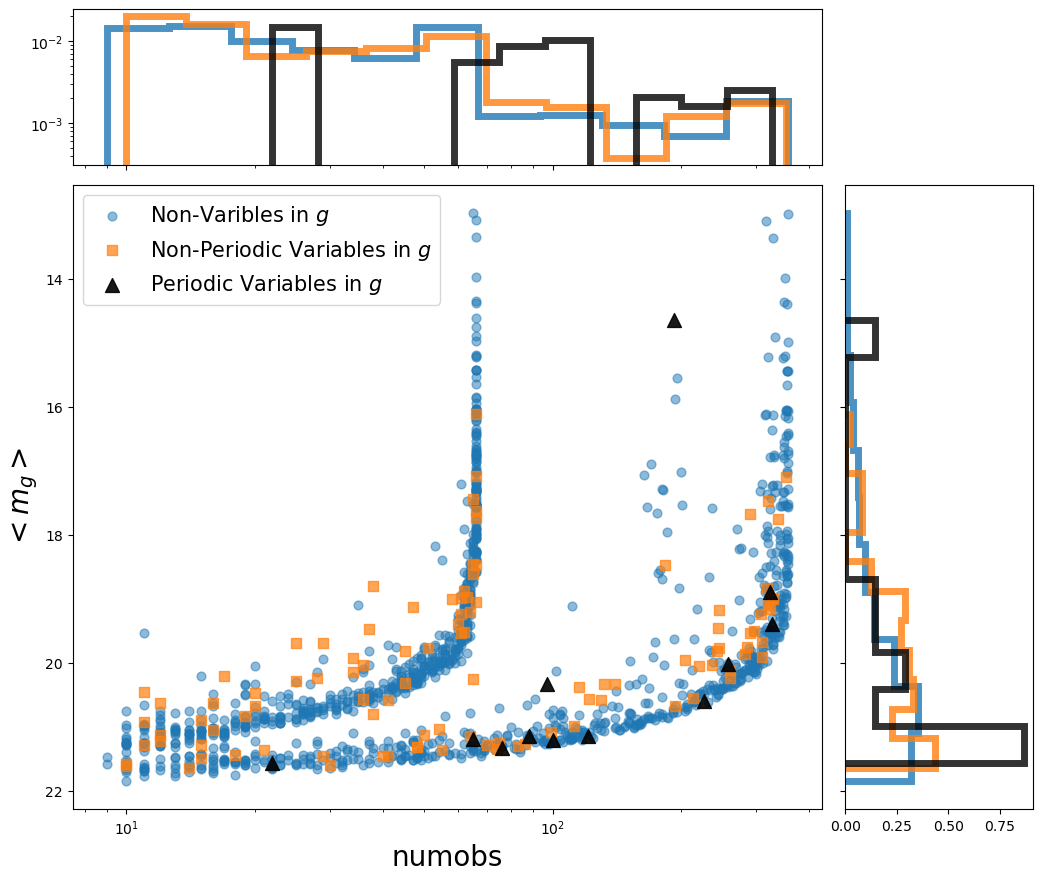}}
\caption{Number of detections verse the mean magnitude of each lightcurve labeled in ZTF OIDs. Top and right panels show the normalized probability density distribution (total area under the curve sums to one) of $x$ and $y$ axis quantities respectively. Objects with lower magnitude has higher numbers of observations simply because brighter sources are more easily detected. However, there are two separate lines of such trend because one real source usually have multiple (two in most cases) ZTF OIDs.}
\label{fig:numobs}
\end{figure*}

\subsection{Classification verification}
\label{sec:veri}

\subsubsection{Cross-check to Chen et al.}
\label{sec:chen}
There are numerous past studies which identify variable stars using ZTF data. \cite{Chen_2020} published a ZTF DR2 catalog of 781,602 periodic variable stars, using Lomb-Scargle periodogram based false-alarm probability to identify these sources, and further classified these periodic variable into 11 class labels according to their distribution across many parameters like period, phase difference, amplitude, amplitude ratio, and absolute magnitude.

\begin{figure}
\centering
\subfigure{
\includegraphics[width=0.9\linewidth]{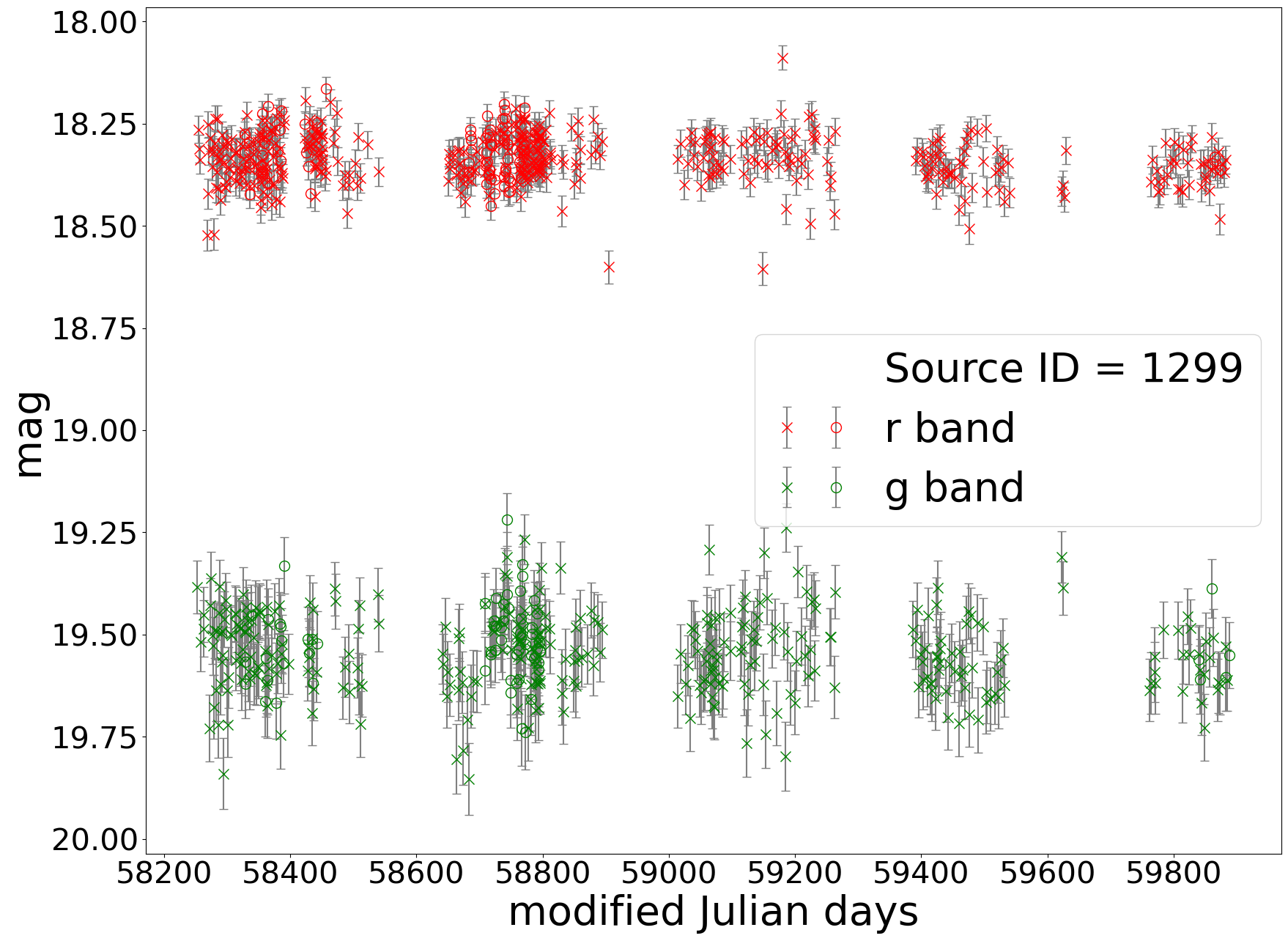}}
\subfigure{
\includegraphics[width=0.9\linewidth]{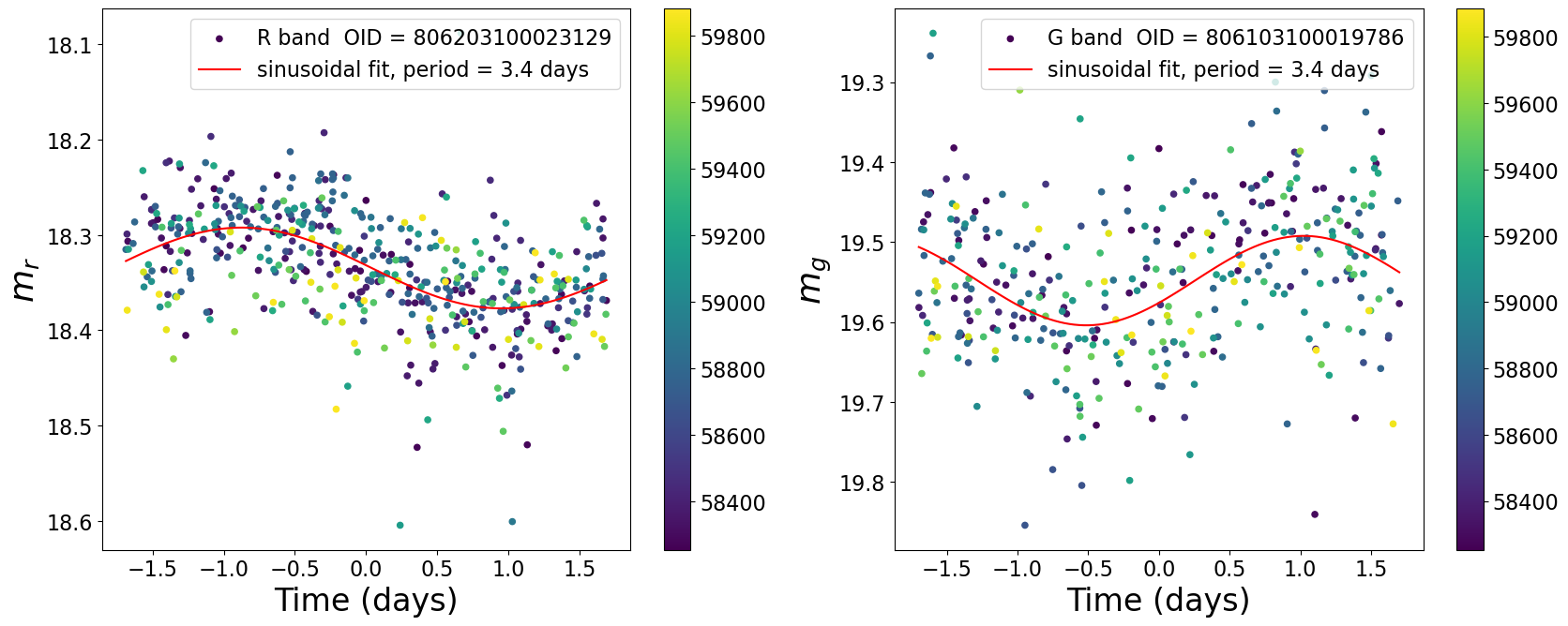}}
\caption{The periodic variable source found by Chen et al. in IC 10. Top panel shows the lightcurve with error in magnitude as error bar. Different marker represents different ZTF labeled OIDs under the same Source ID. Bottom panel are the folded lightcurve of the two OIDs, fitted with a sine function, and color-coded by mjd. This source is identified as periodic variable in $r$ by our algorithm, but did not pass the survival function test in $g$ band due the relatively large error in $g$ band magnitude measurements.}
\label{fig:1299}
\end{figure}

Only one periodic variable star identified by \cite{Chen_2020} falls in the field of IC 10, which is re-discovered in this work. Its lightcurve is shown in Figure \ref{fig:1299}, and identified as a periodic variable in $r$ band, while in $g$ band it does not pass our survival function test. If we only use FAP to determine periodicity, as \cite{Chen_2020} does, it would be a period variable in both bands.
Furthermore, we determine the period of this source is 3.4 days, the same as the period found by \cite{Chen_2020}.

\subsubsection{UPSILoN, a machine learning powered classification}
\label{sec:upsilon}
To cross check our methodology, we searched for periodic variables in our collection of ZTF OIDs using UPSILoN (AUtomated Classification of Periodic Variable Stars using MachIne LearNing), a machine learning driven algorithm to identify and classify periodic variables in any optical survey developed by \cite{2015ascl.soft12019K}. UPSILoN first extracts key features of a lightcurve such as the amplitude, period, FAP, kurtosis, phase difference, and Shapiro-Wilk test statistics, and then a random forest model deploys these features to classify the lightcurve into classes including non-variable, $\delta$ Scuti, RR Lyrae, Cepheid, Type II Cepheid, eclipsing binary, long-period variable, and their respective subclasses. We run UPSILoN on all OIDs within the IC 10 field. The detailed subclasses output of UPSILoN is included in our catalog. 

\begin{figure}
\centering
\includegraphics[width=0.9\linewidth]{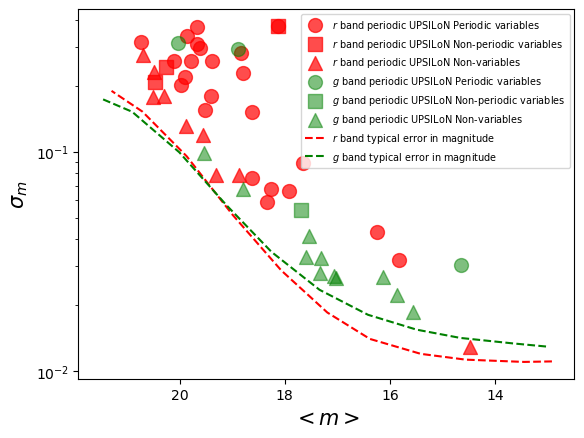}
\caption{The mean of magnitudes and the standard deviation of magnitudes for UPSILoN periodic sources, with shapes indicating the classification in this work. The dashed line shows the typical observation error in magnitudes at certain magnitude. The error in magnitudes is larger for dimmer (larger $<m>$ sources). UPSILoN periodic sources identified as non-variable in this work typically have lower variation (low $<\sigma_m>$) in magnitude compared to variable sources at the similiar $<m>$, and are closer to the dashed line of error in magnitude measurements, showing that our sample of variable sources is prioritizing on purity instead of completeness.}
\label{fig:upsilon_mean_std}
\end{figure}

\begin{figure}
\centering
\subfigure{
\includegraphics[width=0.9\linewidth]{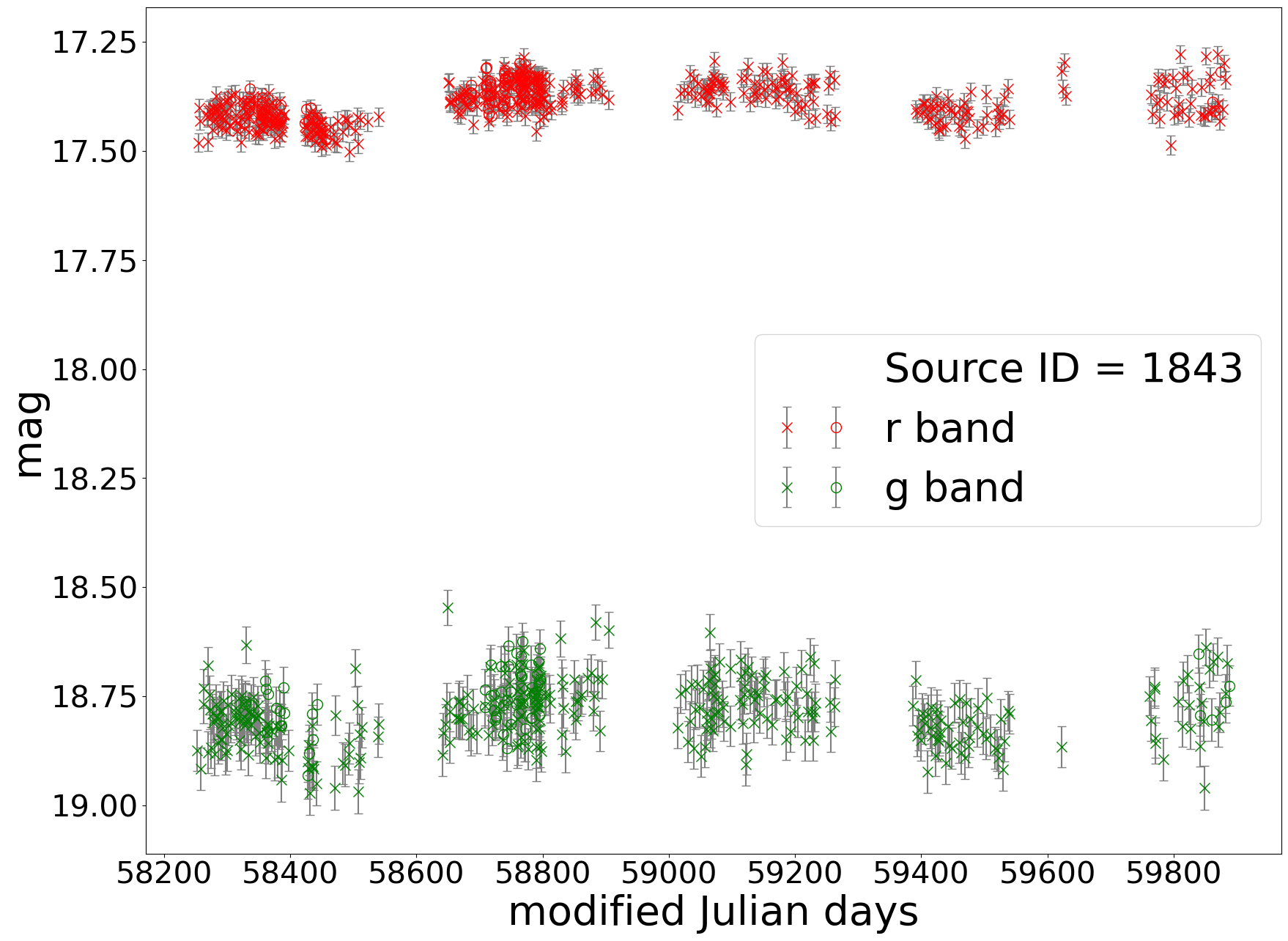}}
\subfigure{
\includegraphics[width=0.9\linewidth]{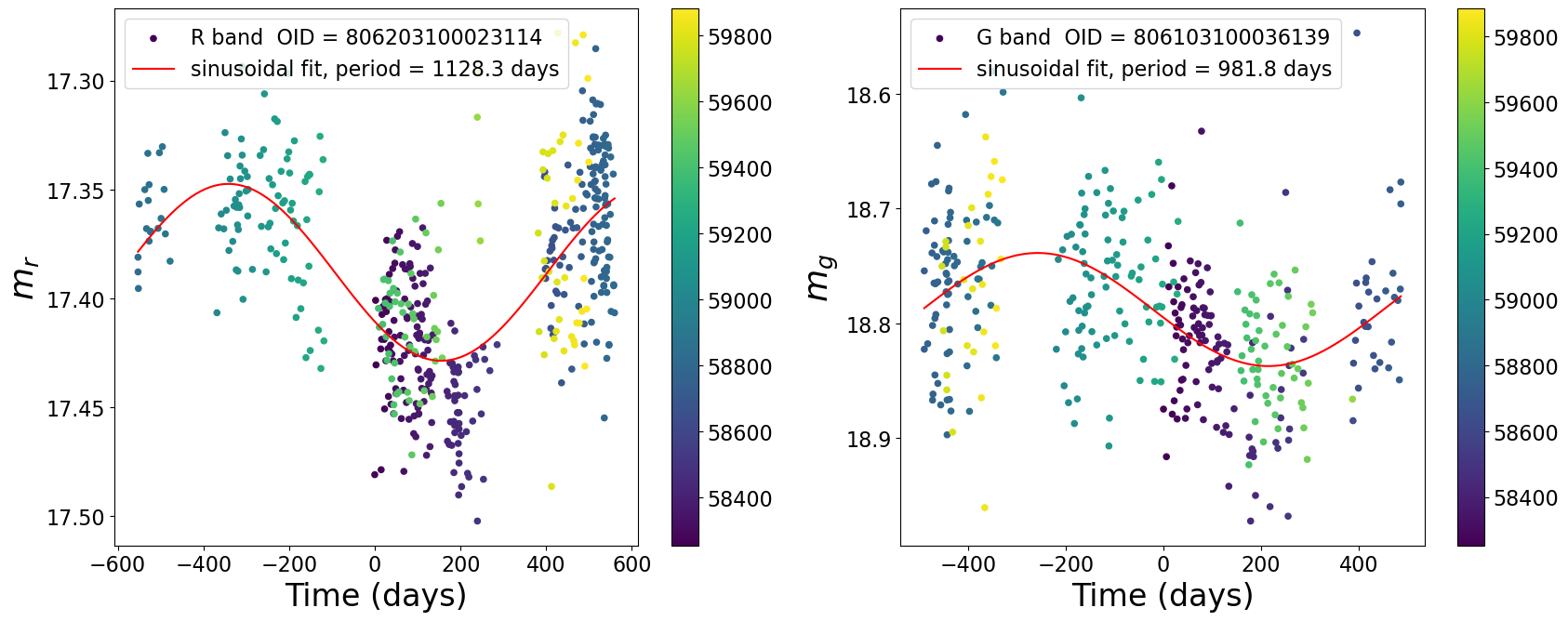}}
\caption{An example source that is identified as periodic in $r$ band by our catalog but not by UPSILoN. Top panel shows the lightcurve with error in magnitude as error bar. Different markers represent two ZTF labeled OIDs under the same Source ID. Bottom panel are the folded lightcurve of the two OIDs, fitted with a sine function, and color-coded by mjd. The FAP of the $r$ band periodicity is $7.29\times10^{-64}$, and the $\chi^2$ of the $r$ band sinusoidal fit is $2.36$.}
\label{fig:439}
\end{figure}

For the purpose of cross-checking, we treat any source (results from multiple lightcurves for a same astronomical sources is grouped the same way as in \S\ref{sec:group}) not classified by UPSILoN as ``non-variable", and not flagged ``suspicious" as a periodic variable identified by this software. 
An overview of the results can be found in Table \ref{tab:upsilon}.
In this work, a source is classified as non-variable only when its lightcurve is consistent with a constant magnitude (see \S\ref{sec:variabilitycut}), while in UPSILoN there is no explicit comparison between the variation in magnitude and the observational error in the apparent magnitude. 
As shown in Figure \ref{fig:upsilon_mean_std}, the rms magnitude of UPSILoN periodic sources are larger than the typical observational error for their $<m>$, but those classified as “non-variable” in this work are lower than those we classify as “variable.”  
The “variable” UPSILoN periodic sources are largely classified as periodic in this work (23/26 in r, 3/4 in g), with the two measured periods agreeing to better than 10\%.
On the other hand, we identified 14 and 9 periodic variables in $r$ band and $g$ band missed by UPSILoN.
Figure \ref{fig:439} is a typical example where we identify a source as periodic variable but UPSILoN did not. This source definitely looks variable from its lightcurve, and shows some periodicity at around 238 days from its folded lightcurve.

\begin{table*}[]
\begin{tabular}{ccccc}
\multirow{2}{*}{} & \multirow{2}{*}{UPSILoN Periodic variables} & \multicolumn{3}{c}{Classification in this work}                                        \\ %
                  &                                             & \multicolumn{1}{c}{Periodic variable} & \multicolumn{1}{c}{Non-periodic Variable} & Non-variable \\ \hline
r band            & 35                                          & \multicolumn{1}{c}{23}                & \multicolumn{1}{c}{3}                     & 9            \\ \hline
g band            & 15                                          & \multicolumn{1}{c}{3}                 & \multicolumn{1}{c}{1}                     & 11           \\ 
\end{tabular}
\caption{\label{tab:upsilon}Periodic sources found by machine-learning-driven UPSILoN and the classification of those sources in this work.}
\end{table*}

\subsection{Catalog columns}
\label{sec:columns}
All columns of our catalog are presented in Table \ref{tab:columns}. The catalog itself, alone with the code to reproduce all the plots in this paper can be downloaded at \url{https://github.com/ZehaoJin/Transients-in-IC-10}. 

\begin{table*}
\centering
\begin{tabular}{ll}
\texttt{SourceID} & Source ID for this lightcurve/OID. OIDs with less than $1''$ have the same \texttt{SourceID}.  \\ \hline
\texttt{filter} & ZTF filter, either zr for $r$ band or zg for $g$ band. \\ \hline
\texttt{OID} & ZTF Object ID. Each OID has a lightcurve and makes up a row in this paper's catalog.  \\ \hline
\texttt{RA} & Right ascension, from ZTF.  \\ \hline
\texttt{DEC} & Declination, from ZTF.  \\ \hline
\texttt{numobs} & Number of observations by ZTF, or number of points of this lightcurve. \\ \hline
\texttt{sf} & $S_k(\chi^2)$, survival function of the constant magnitude model, to classify variable vs. non-variable.  \\ \hline
\texttt{FAP} & False alarm probability from Lomb-Scargle, to classify periodic variable vs. non-periodic variable.\\ \hline
\texttt{period} & $T_{\rm LS}$, period at Lomb-Scargle periodogram peak power. Period of the source if the source is periodic.  \\ \hline
\texttt{period$\_$w} & $T_{\rm window}$, period at window function peak power. $C_{T}=|T_{\rm LS}-T_{\rm window}|/T_{\rm window}$ for de-aliasing. \\ \hline
\texttt{power} & $P_{\rm LS}$, Lomb-Scargle periodogram peak power.  \\ \hline
\texttt{power$\_$w} & $P_{\rm window}$, window function power at $T_{\rm LS}$. $Q_{P}=P_{\rm LS}/P_{\rm window}$ for de-aliasing.  \\ \hline
\texttt{mean} & Mean magnitude.  \\ \hline
\texttt{std} & Standard deviation of magnitudes.  \\ \hline
\texttt{min} & Minimum magnitude.  \\ \hline
\texttt{max} & Maximum magnitude.  \\ \hline
\texttt{upsilon$\_$class} & Classification label from UPSILoN.  \\ \hline
\texttt{upsilon$\_$prob} & Classification probability from UPSILoN.  \\ \hline
\texttt{upsilon$\_$flag} & Suspicious classification flag by UPSILoN.  \\ \hline
\texttt{upsilon$\_$period} & Period found by UPSILoN.  \\ \hline
\texttt{identifier} & SIMBAD database identifier of the nearest (if there is any) match within $2''$. \\ \hline
\texttt{dist} & Angular distance between ZTF location and SIMBAD location of the nearest match. \\ \hline
\texttt{type} & SIMBAD object type of the nearest match. \\ \hline
\texttt{Gaia$\_$count} & Number of Gaia objects within $2''$ radius.  \\ \hline
\texttt{multiflag} & If there is possible contamination from multiple unresolved sources, see \S \ref{sec:gaia}.  \\ \hline
\texttt{GaiaDist} & Angular distance between ZTF location and Gaia location of the brightest match.  \\ \hline
\texttt{fluxratio} & Gaia \texttt{G} band flux of brightest Gaia match over that of second brightest Gaia match.  \\ \hline
\texttt{Gmag} & Gaia \texttt{G} band magnitude.  \\ \hline
\texttt{Plx} & Parallax of the brightest Gaia match.  \\ \hline
\texttt{e$\_$Plx} & Error in parallax of the brightest Gaia match.  \\ \hline
\texttt{Plxflag} & If the source is outside IC 10.  \\ \hline
\texttt{RUWE} & Renormalised Unit Weight Error (RUWE) of the brightest Gaia match.  \\ \hline
\texttt{XMM} & Number of XMM objects. \\ \hline
\texttt{XMM$\_$dist} & Angular distance between between ZTF location and XMM location \\ \hline
\texttt{Chandra} & Number of Chandra objects. \\ \hline
\texttt{Chandra$\_$dist} & Angular distance between between ZTF location and Chandra location \\ \hline
\texttt{inside$\_$flag} & If the source is within the $25\sigma$ overdensed region. See \S\ref{sec:defineic10}. \\ \hline
\texttt{ic10$\_$flag} & If the source is likely to be inside IC 10, with \texttt{Plxflag}=0 and \texttt{inside$\_$flag}=1. See \S\ref{sec:defineic10}. \\ \hline
\texttt{lc$\_$var} & Classification of this lightcurve/OID. \texttt{P} = periodic variable, \texttt{V} = non-periodic variable, \texttt{N} = non-variable. \\ \hline
\texttt{var} &  Classification of this source, \texttt{P/V/N}.  \\ \hline
\texttt{var$\_$flag} & Variable suspect flag, see \S \ref{sec:group}.  \\ \hline
\end{tabular}
\caption{\label{tab:columns}Columns of the catalog.}
\end{table*}

\section{Catalog demographics}
\label{sec:demo}

\subsection{General demographics}
A summary of the number of sources (as described in \S \ref{sec:group}) classified as variables, periodic variables, and non-periodic variables can be found in Table \ref{tab:demo}. 
There are more sources observed, and exclusively observed in $r$ band than $g$ band because sources in this field are typically $\sim 1$ magnitude ($\sim 2.5$ times) brighter in $r$ band than in $g$ band (Fig.~\ref{fig:cmd}), making $r$ band relatively more sensitive. 
In both bands, $\sim10\%$ of the sources are variable (periodic variables \& non-periodic variables) in one of these bands, but only $\sim4\%$ are variable in both. 
Furthermore, in each band only $\sim1-2\%$ are periodic.

\begin{table*}
\begin{tabular}{llllll}
                       & $r$ band & $g$ band & Both $r$ and $g$ & Only detected in $r$ & Only detected in $g$\\ \hline
All Sources            & 1516 (100\%)   & 864 (100\%)    & 821 (100\%)   & 695 (100\%)  & 43 (100\%)     \\ \hline
Non-Variables          & 1329 (87.7\%)  & 767 (88.8\%)   & 681 (82.9\%)  & 611 (87.9\%)  & 25 (58.1\%)      \\ \hline
Periodic Variables     & 37 (2.4\%)     & 12 (1.4\%)     & 2 (0.2\%)     & 18 (2.6\%)   & 0 (0\%)       \\ \hline
Non-Periodic Variables & 150 (9.9\%)    & 85 (9.8\%)     & 34 (4.1\%)    & 66 (9.5\%)   & 18 (41.9\%)      
\end{tabular}
\caption{\label{tab:demo}General statistics of the catalog. Column name ``$r$/$g$ band" means the sources are observed in $r$/$g$ band, and is variable/periodic/non-periodic in $r$/$g$ band; ``Both $r$ and $g$" means the sources are observed in both $r$ and $g$ band, and is variable/periodic/non-periodic in both $r$ and $g$ band; ``Only detected in $r$/$g$" means the sources are only observed in $r$/$g$ band, never seen in the other($g$/$r$) band, and is variable/periodic/non-periodic in $r$/$g$ band. The percentages mark the percentage inside each column.}
\end{table*}

\subsection{Color-magnitude Diagrams}
\begin{figure}
\centering
\includegraphics[width=0.9\linewidth]{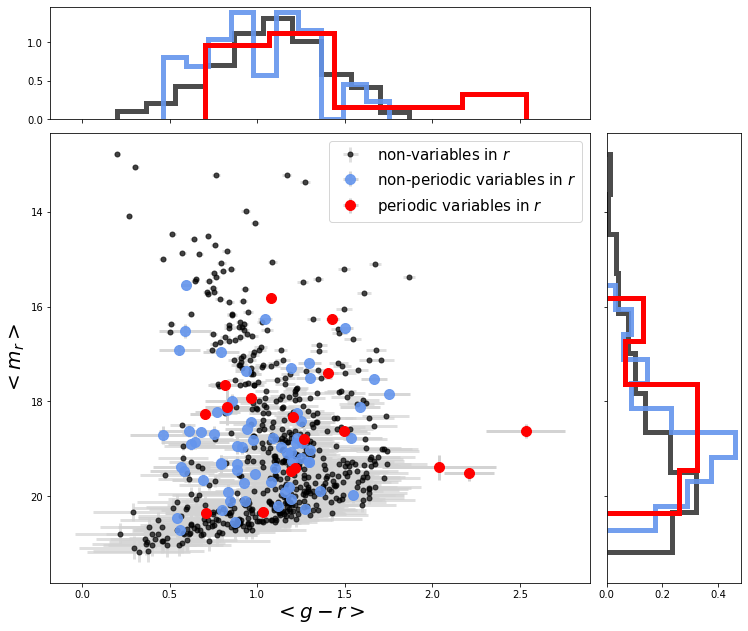}
\caption{The color magnitude diagram. On horizontal axis is the mean $g$-$r$ color, and on vertical axis is the mean $r$ magnitude. Grey lines denotes $1\sigma$ in color and magnitude. Top and right panel shows the normalized probability distribution of $x$ and $y$ axis quantities respectively. Only sources detected in both $r$ and $g$ band are included in this plot.}
\label{fig:cmd}
\end{figure}

Figure \ref{fig:cmd} shows the magnitude and color distribution of sources detected in both bands. 
Variable sources are well separated in color and brightness.
In terms of average magnitude, variable sources are slightly brighter than non-variable ones, since the lower uncertainty in the magnitude lowers the threshold needed to detected variability.
In terms of colors, non-periodic variables tends to be bluer (smaller $g$-$r$), and periodic variables are redder than non-variable objects. 
The three very red periodic sources on the right part of the plot will be further discussed in \S \ref{sec:superred}.

\subsection{Counterparts in SIMBAD, Chandra and XMM}
\label{sec:simbadmatch}
SIMBAD, or Set of Identifications, Measurements, and Bibliography for Astronomical Data, is a comprehensive online database that provides information on celestial objects such as stars, galaxies, and planetary nebulae. Developed and maintained by the Centre de Données astronomiques de Strasbourg (CDS) in France, SIMBAD is a widely-used tool for astronomers and astrophysicists to access data related to various celestial objects. 

\begin{figure}
\centering
\includegraphics[width=0.9\linewidth]{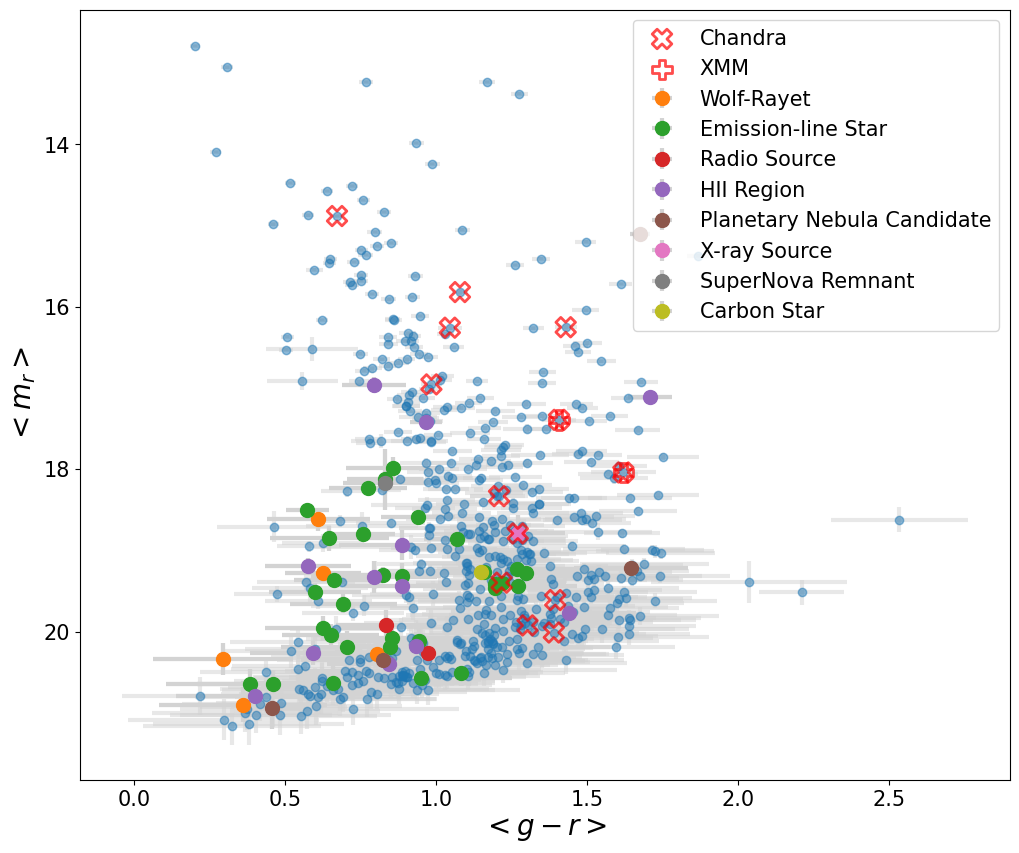}
\caption{Sources labeled by SIMBAD on the color magnitude diagram. Blue dots are all the sources included in this catalog, colored dots are SIMBAD matches of different object types, and Grey lines denotes $1\sigma$ in color and magnitude. \emph{XMM-Newton} and \emph{Chandra} counterparts are labeled as crosses ($\times$) and pluses ($+$).}
\label{fig:simbad}
\end{figure}

We queried the SIMBAD database for sources within $2^{\prime \prime}$ of each ZTF source, considering the nearest (if any) object as its matched counterparts. An overview of the matched statistics is shown in Table \ref{tab:simbad}, the detailed matched information including identifier name, identified type, and distance is include in our catalog. Among our ZTF sources, SIMBAD indicates the presence of Wolf-Rayet stars, emission-line stars, radio sources, HII regions, candidate planetary nebula, X-ray sources, supernova remnants and carbon stars. For those detected in both ZTF $r$ and $g$ band, their distribution on a color magnitude diagram is shown in Figure \ref{fig:simbad}. Objects of the same type usually share similar physics and thus tends to cluster at certain places in the color-magnitude diagram. Table \ref{tab:simbadclass} details the variability of SIMBAD identified sources. The variability rate can serve as a rough guideline for the those previously not identified sources.

Both \emph{Chandra} and \emph{XMM-Newton} have both surveyed the field of IC~10, identifying a number of X-ray sources. We match our catalog with the IC 10 \emph{Chandra} X-Ray point source catalog \citep{2017ApJ...836...50L}, and the IC 10 \emph{XMM-Newton} X-Ray point source catalog \citep{2005MNRAS.362.1065W}, using a $2^{\prime\prime}$ search radius. Their location on a color diagram is shown in Figure \ref{fig:simbad}. We find two non-periodic variable ZTF sources coincident with a \emph{Chandra}  X-ray source , six periodically variable ZTF sources with counterparts in the \emph{Chandra} catalog, one of the six also has a match in the \emph{XMM} catalog. The period of these X-ray sources ranges from 3.4 days to 1608 days. One of the latest Galactic X-Ray Binary catalog XRBcats \citep{Avakyan_2023,Neumann_2023} finds low-mass X-ray binaries (LMXBs) with periods from 0.2-300 days, with a couple at $\sim$1200 days, and high mass X-ray binaries (HMXBs) with periods from much less than 1 day to around 30 days, as well as some at thousands days.

\begin{table}
\resizebox{\linewidth}{!}{
\begin{tabular}{lll}
                            & This work & Simbad matches \\ \hline
Periodic variables in $r$     & 37        & 15             \\ \hline
Periodic variables in $g$     & 12         & 4              \\ \hline
Non-Periodic variables in $r$ & 150       & 42             \\ \hline
Non-Periodic variables in $g$ & 85        & 25            
\end{tabular}
}
\caption{\label{tab:simbad}Number of SIMBAD matches within 2 arcsec radius.}
\end{table}

\begin{table*}
\resizebox{\linewidth}{!}{
\begin{tabular}{lllllllll}
                           & \multicolumn{2}{l}{Total Matches}   & \multicolumn{2}{l}{Periodic Variable} & \multicolumn{2}{l}{Non-Periodic Variable} & \multicolumn{2}{l}{Non-Variable} \\ \cline{1-9} 
                           & \multicolumn{1}{l}{$r$}  & $g$  & \multicolumn{1}{l}{$r$}       & $g$       & \multicolumn{1}{l}{$r$}         & $g$         & \multicolumn{1}{l}{$r$}    & $g$    \\ \hline
Wolf-Rayet                 & \multicolumn{1}{l}{18 (100\%)} & 16 (100\%) & \multicolumn{1}{l}{0 (0\%)}       & 0 (0\%)      & \multicolumn{1}{l}{4 (22.2\%)}         & 4 (25\%)         & \multicolumn{1}{l}{14 (77.8\%)}   & 12 (75\%)   \\ \hline
Emission-line Star         & \multicolumn{1}{l}{50 (100\%)} & 40 (100\%) & \multicolumn{1}{l}{3 (6\%)}       & 1 (2.5\%)       & \multicolumn{1}{l}{9 (18\%)}         & 7 (17.5\%)         & \multicolumn{1}{l}{38 (76\%)}   & 32 (80\%)   \\ \hline
Radio Source               & \multicolumn{1}{l}{6 (100\%)}  & 2 (100\%)  & \multicolumn{1}{l}{0 (0\%)}       & 0 (0\%)       & \multicolumn{1}{l}{3 (50\%)}         & 0 (0\%)         & \multicolumn{1}{l}{3 (50\%)}    & 2 (100\%)    \\ \hline
HII Region                 & \multicolumn{1}{l}{33 (100\%)} & 16 (100\%) & \multicolumn{1}{l}{3 (9.1\%)}       & 1 (6.3\%)       & \multicolumn{1}{l}{5 (15.2\%)}         & 5 (31.2\%)         & \multicolumn{1}{l}{25 (75.8\%)}   & 10 (62.5\%)   \\ \hline
Planetary Nebula Candidate & \multicolumn{1}{l}{8 (100\%)}  & 6 (100\%)  & \multicolumn{1}{l}{0 (0\%)}       & 0 (0\%)       & \multicolumn{1}{l}{0 (0\%)}         & 0 (0\%)         & \multicolumn{1}{l}{8 (100\%)}    & 6 (100\%)    \\ \hline
X-ray Source               & \multicolumn{1}{l}{1 (100\%)}  & 1 (100\%)  & \multicolumn{1}{l}{1 (100\%)}       & 1 (100\%)       & \multicolumn{1}{l}{0 (0\%)}         & 0 (0\%)         & \multicolumn{1}{l}{0 (0\%)}    & 0 (0\%)    \\ \hline
SuperNova Remnant          & \multicolumn{1}{l}{1 (100\%)}  & 3 (100\%)  & \multicolumn{1}{l}{0 (0\%)}       & 0 (0\%)       & \multicolumn{1}{l}{1 (100\%)}         & 2 (66.7\%)         & \multicolumn{1}{l}{0 (0\%)}    & 1 (33.3\%)    \\ \hline
Carbon Star                & \multicolumn{1}{l}{14 (100\%)} & 5 (100\%)  & \multicolumn{1}{l}{1 (7.1\%)}       & 0 (0\%)       & \multicolumn{1}{l}{0 (0\%)}         & 1 (20\%)         & \multicolumn{1}{l}{13 (92.9\%)}   & 4 (80\%)  
\end{tabular}
}
\caption{\label{tab:simbadclass}Classification statistics for SIMBAD matches. The percentages show the percentage in each row, with $r$ and $g$ band separated.}
\end{table*}

\subsection{Distinguish foreground and distant sources with Gaia}
\label{sec:gaia}
Gaia is launched by the European Space Agency (ESA) with the goal of creating a highly precise three-dimensional map of our Milky Way galaxy. We cross match sources in our (ZTF) catalog with Gaia DR3 sources using a $2''$ radius distance. A summary is shown in Table \ref{tab:Gaia}. 

Thanks to Gaia's higher angular resolution ($\sim 0.4''$,\cite{gaiadr2,gaiadr3}), this cross-matching to Gaia will tell us how many resolved real sources are under our same source label and avoid possible contamination. For ZTF sources with more than one Gaia matches within $2''$, if the brightest (in terms of Gaia \texttt{G} band magnitude) Gaia match is not 100 times brighter (2.5 magnitudes lower) than the second brightest Gaia match, we place \texttt{multiflag=1} in the catalog to indicate possible contamination from multiple unresolved sources.

\begin{figure}
\centering
\includegraphics[width=0.9\linewidth]{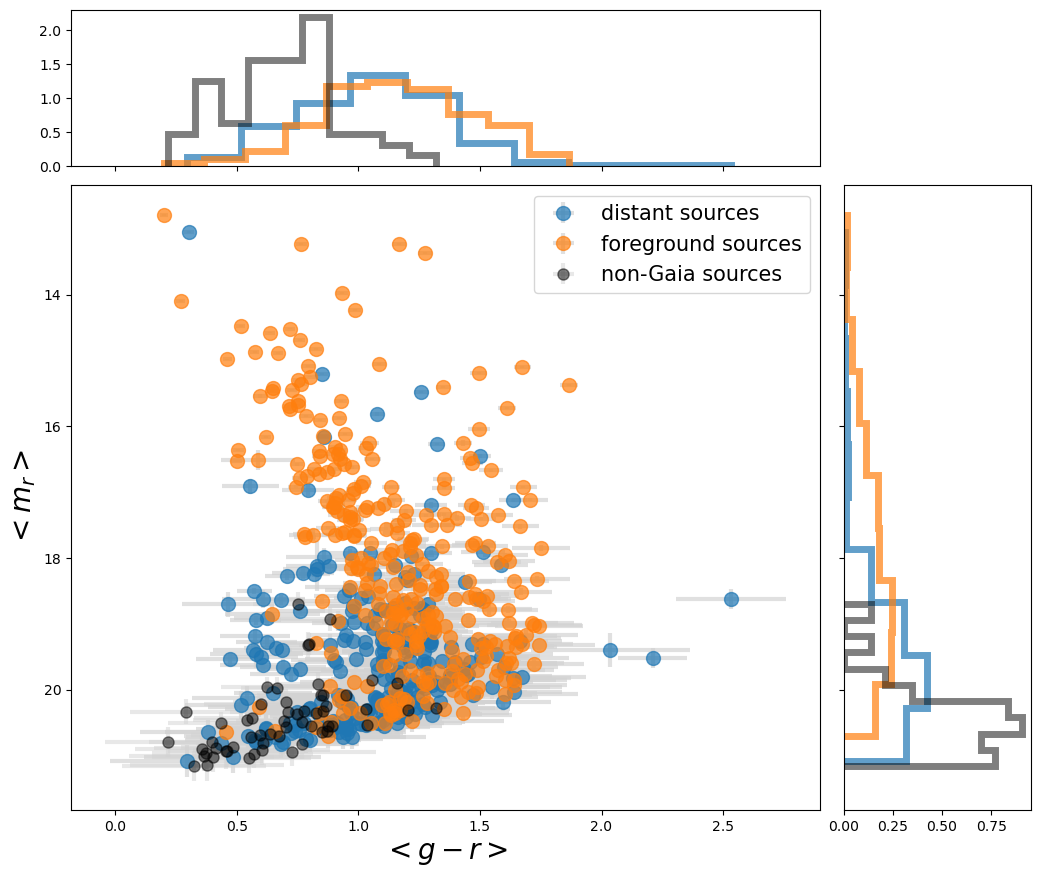}
\caption{The distribution of foreground sources (\texttt{Plxflag=1}), distant sources (\texttt{Plxflag=0}), and non-Gaia sources on a color magnitude diagram. Top and right panel shows the normalized distribution of $x$ and $y$ axis quantities respectively. Foreground sources are generally brighter and redder (higher $g$-$r$ color).}
\label{fig:plx}
\end{figure}

IC 10 is outside of Milky Way, and is around 950 kpc away \citep{1995AJ....109.2470M}. Gaia measures the parallax, an indicator of how far the source is away from us. Therefore, we can distinguish nearby sources inside Milky Way (MW) and the further non-MW sources, probably within IC 10 galaxy. The smallest median parallax uncertainties are 0.02-0.03 mas (corresponds to 50-33 kpc) for Gaia $m_g<15$ \citep{Gaia_2023}, therefore parallax measurements are highly impossible for IC 10 sources which are 950 kpc away. Therefore we place a \texttt{Plxflag=1} for sources with a meaningful parallax (SNR\footnote{Here we adopt a relatively loose cut on parallax SNR (SNR$>$1), because we are mainly interested in IC 10 sources which will have insignificant SNR, but not MW sources with significant SNR. A source with more significant SNR (e.g. SNR$>$3) is very likely to be a MW source, a source with a relatively lower SNR (e.g. 1$<$SNR$<$3) is possibly a MW source, and a source with insignificant SNR (e.g. SNR$<$1) is more likely to not be in the MW, and therefore potentially be in IC 10.}, i.e. abs(Plx/e\_Plx)$>$1, and RUWE $<$ 1.4 \citep{Lindegren_2018,Lindegren_2021}) to indicate that these sources are likely foreground to IC 10. Figure \ref{fig:plx} puts foreground source with \texttt{Plxflag=1} and distant source with \texttt{Plxflag=0} on a color magnitude diagram. Not surprisingly, Foreground sources appears to be brighter than distant sources. In addition, foreground sources have higher $g$-$r$ colors, in other words, redder than distant sources. This suggests that these further sources, which are most likely sources within the "star-bursting" galaxy IC 10, are younger than nearby sources. It is also worth to note that on the lower left region of the color magnitude diagram ($<m_r>$ around $18-20$, and $<g-r>$ around $0.4-1.0$) there is a region where most of the sources are distant sources. It is reasonable to expect sources without Gaia matches that falls into this region have a good chance to be also distant sources. 

\begin{figure}
\centering
\includegraphics[width=0.9\linewidth]{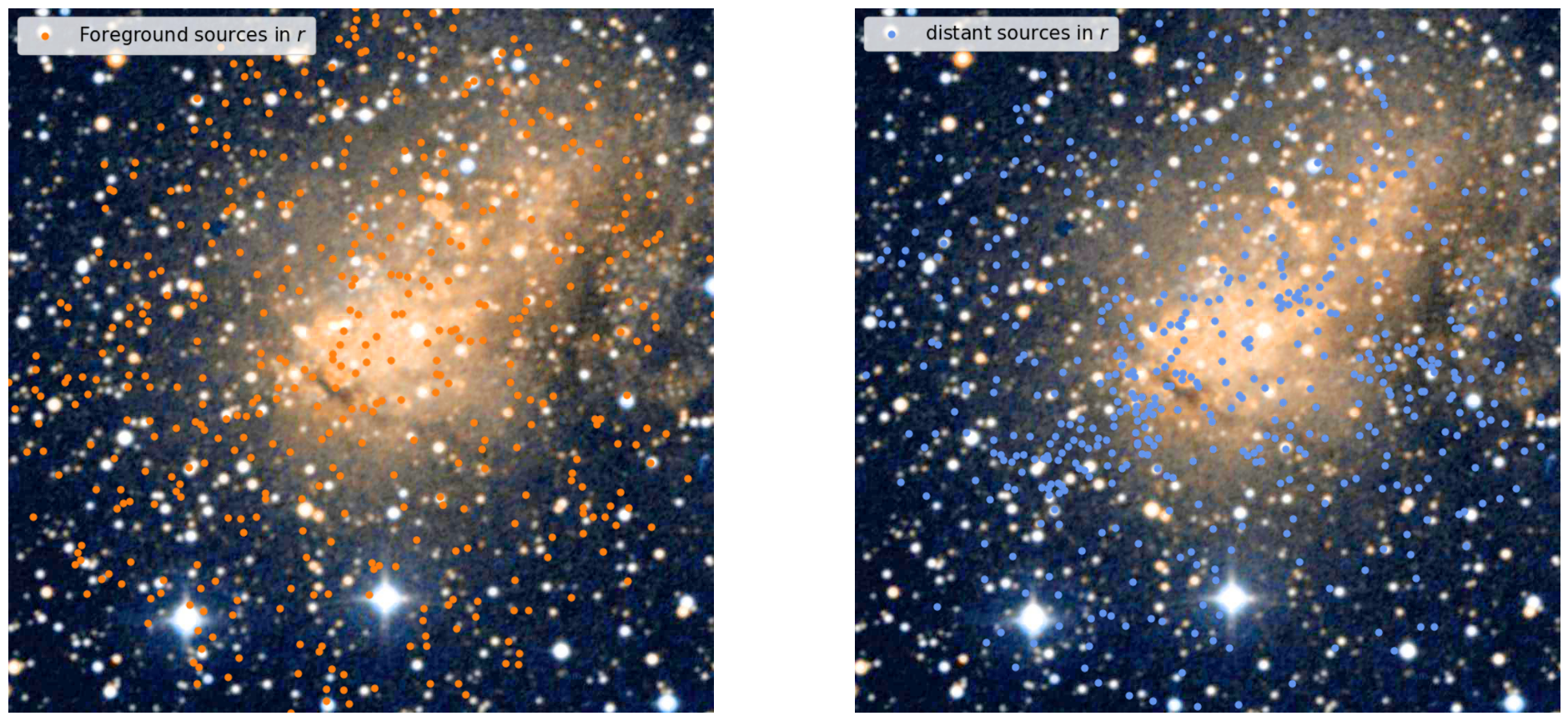}
\caption{Physical location of $r$ band foreground sources (left) and distant (middle) sources, overlaid with a Digital Sky Survey (DSS) color image of IC 10. Foreground sources are more uniformly distributed than distant sources.}
\label{fig:overlay_foreground}
\end{figure}

Figure \ref{fig:overlay_foreground} shows the location of foreground sources and distant sources with respect to a Digital Sky Survey (DSS) color image of IC 10. Foreground sources appears to be uniformly distributed over the field, while distant sources show obvious clustering at region of IC 10, indicating that those distant sources are very likely to be sources intrinsic of IC 10.

\begin{table}
\resizebox{\linewidth}{!}{
\begin{tabular}{lllll}
                              & This work & Gaia matches   & multiflag & Distant sources from parallax \\ \hline
Non-variables in $r$          & 1329      & 851            & 41        & 446            \\ \hline
Non-variables in $g$          & 767       & 659            & 47        & 297            \\ \hline
Non-Periodic variables in $r$ & 150       & 96             & 27        & 60             \\ \hline
Non-Periodic variables in $g$ & 85        & 64             & 14        & 39             \\ \hline
Periodic variables in $r$     & 37        & 32             & 6         & 22             \\ \hline
Periodic variables in $g$     & 12        & 12             & 4         & 6              
\end{tabular}
}
\caption{\label{tab:Gaia}Number of Gaia matches within 2 arcsec radius.}
\end{table}

\subsection{Definition of IC 10 sources}
\label{sec:defineic10}
\begin{figure}
\centering
\includegraphics[width=0.9\linewidth]{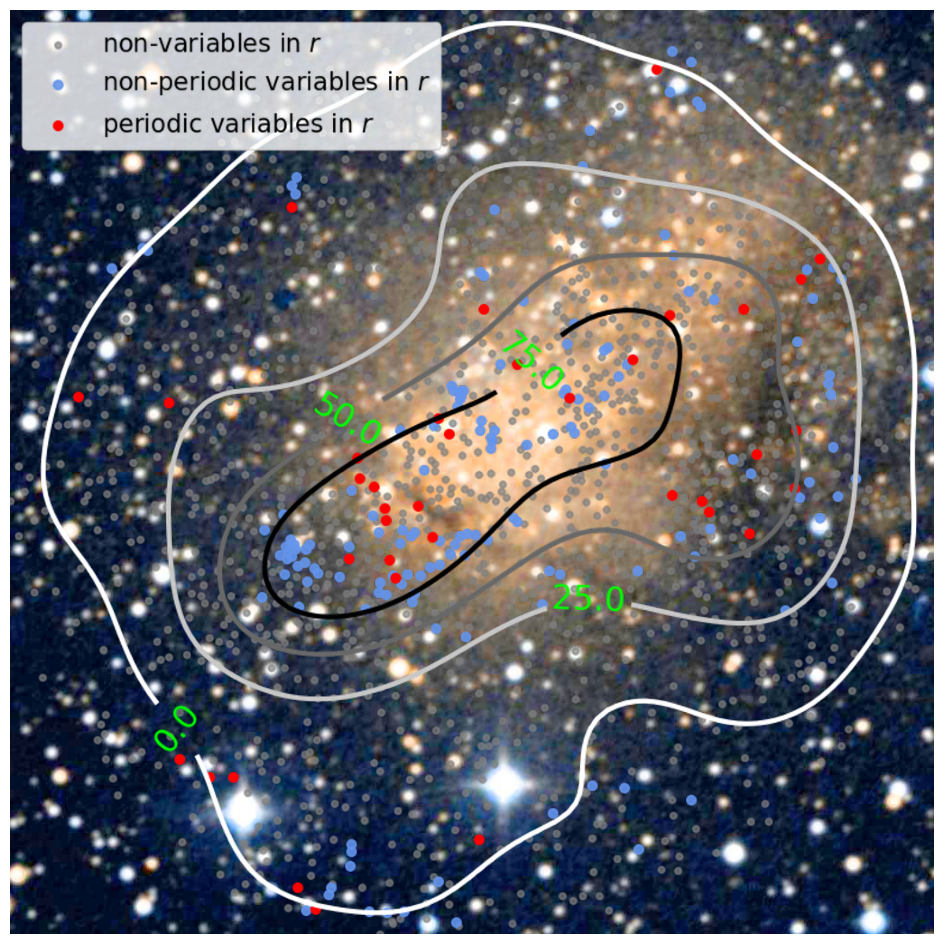}
\caption{Location of ZTF sources overlaid with a Digital Sky Survey (DSS) color image of IC 10. The contours show the source density relative to the mean density of four background fields. The numbers on the contours shows how many sigmas (standard deviation of four background densities) away from mean background density. Sources that are within the 25 $\sigma$ density contour, and are not Gaia foreground sources are defined to be ``IC 10 sources''.}
\label{fig:overlay_density}
\end{figure}

A source on the sky within the IC 10 field would either be a foreground source, a background sources or actually a source within IC 10. Foreground sources and backgrounds are expected to distribute uniformly across the sky, and true IC 10 sources will distribute un-uniformly with overdensity around the center of IC 10. Therefore the overdensed region in the IC 10 field is more likely to host true IC 10 sources. To define such an overdensed region, we compare the source density within IC 10 field to the source density within nearby background fields without the presence of IC 10 galaxy. We look at four regions $675''$ (3 times the IC 10 field radius) above, below, left and right to the IC 10 field with the same ($225''$) search radius as IC 10 field. There are 1044, 1064, 979, and 1059 ZTF sources in the four background fields, with a mean of 1036.5 sources and a standard deviation of 34 sources, while the IC 10 field has 1515 sources in total. In Figure \ref{fig:overlay_density} the contours shows the IC 10 field source density normalized by the mean source density of background field and the standard deviation of background source density, i.e. a contour of 0 means the same density as mean background density, while a contour of 25 means 25 standard deviations higher than the mean background density. To ensure the purity of our sample, we pick a relatively smooth region at 25 $\sigma$ as the IC 10 region, such that sources within this region (with \texttt{inside$\_$flag}=1) of the sky has a higher chance of being inside of IC 10 than sources outside of this region (with \texttt{inside$\_$flag}=0). For consistency, this position cut in $r$ and $g$ band uses the same 25 $\sigma$ density contour calculated with $r$ band source density. For sources within the overdensed IC 10 region, we further cut out those Gaia foreground sources parallax, to arrive at a sample of "IC 10 sources" (\texttt{ic10$\_$flag}=1) that has a good chance of actually inside the IC 10 galaxy. This results in 716 IC 10 sources out of 1516 $r$ band sources, and 337 IC 10 sources out of 864 $g$ band sources.

\section{IC 10 sources}
\label{sec:ic10sources}

\subsection{H-R diagram}
\label{sec:ic10hr}
\begin{figure}
\centering
\includegraphics[width=0.9\linewidth]{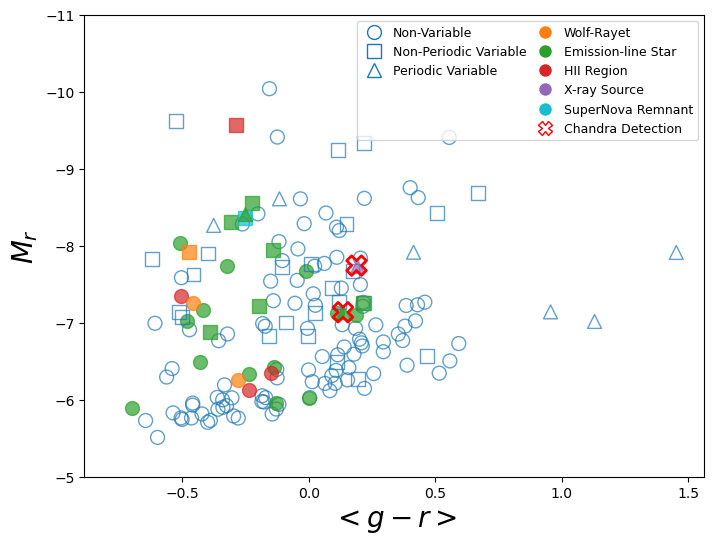}
\caption{H-R diagram for IC 10 sources as defined in \S\ref{sec:defineic10}. The marker shape shows the variability, the color code denotes SIMBAD identification, and the red crosses indicate Chandra detection. There is a non-variable source with $M_r\sim-13$ not drawn in the plot for a better view.}
\label{fig:HRdiagram1}
\end{figure}

For IC 10 sources defined in \S\ref{sec:defineic10}, we adopt the IC 10 distance modulus (m-M) of 24 \citep{Sakai_1999,2005ARBl...20...85O,Kim_2009}, and correct for extinction and reddening towards IC 10 with a total reddening of E(B-V)=0.98 \citep{Kim_2009}, with $A_g/A_v=1.19$ and $A_r/A_v=0.834$ from \cite{Rodrigo_2020} to predict their absolute magnitude. With absolute magnitude and color, a Hertzsprung–Russell (H-R) diagram can be constructed, as shown in Figure \ref{fig:HRdiagram1}. IC 10 sources included in this work are located in giant or super giant branch.

\subsection{Non-periodic variable sources}
\label{sec:ic10-flares}

\begin{figure}
\centering
\includegraphics[width=0.9\linewidth]{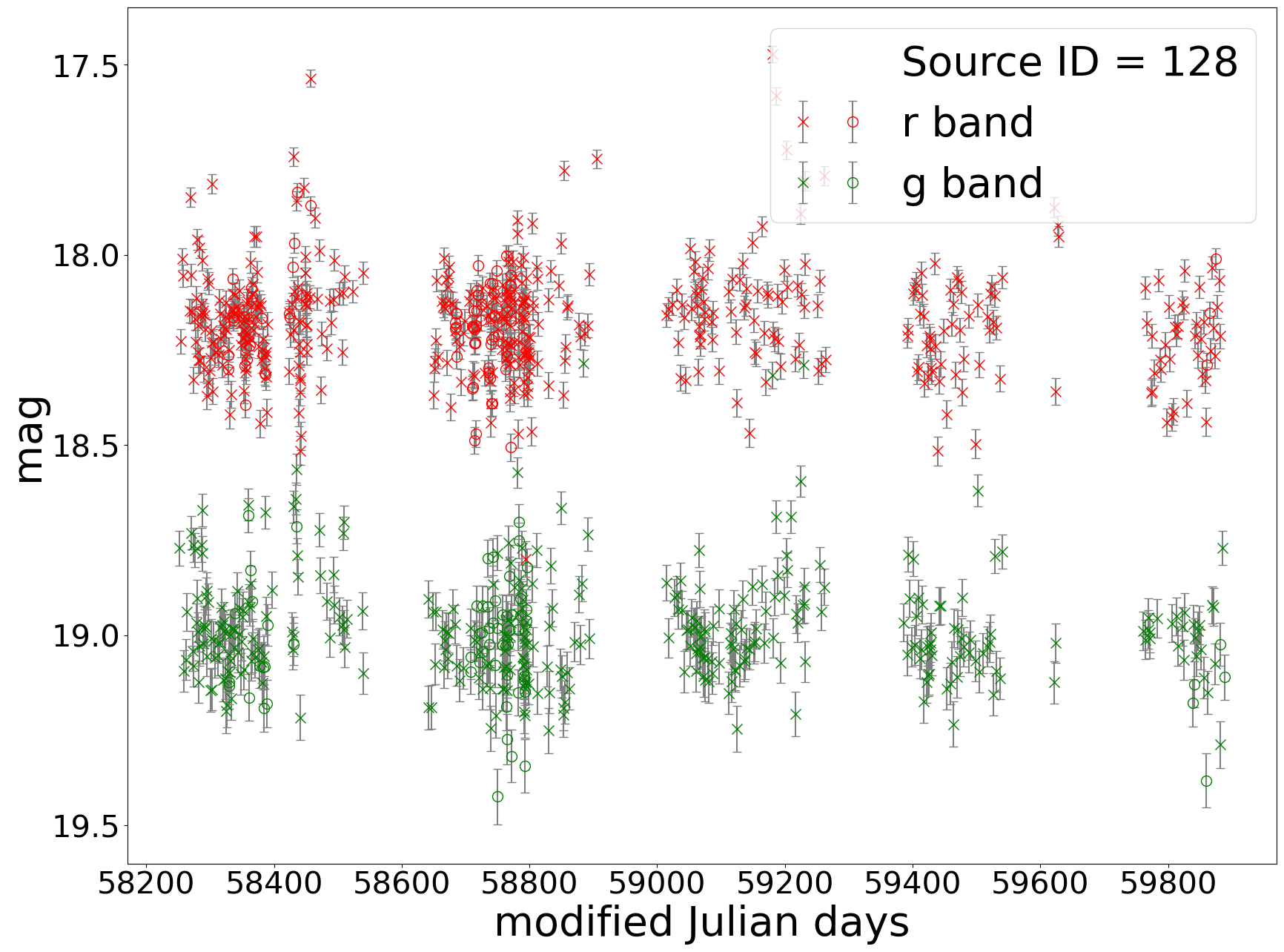}
\caption{Source 128. Different markers represents different ZTF labeled OIDs.}
\label{fig:lc_128}
\end{figure}

\begin{figure}[tbh]
\centering
\includegraphics[width=0.9\linewidth]{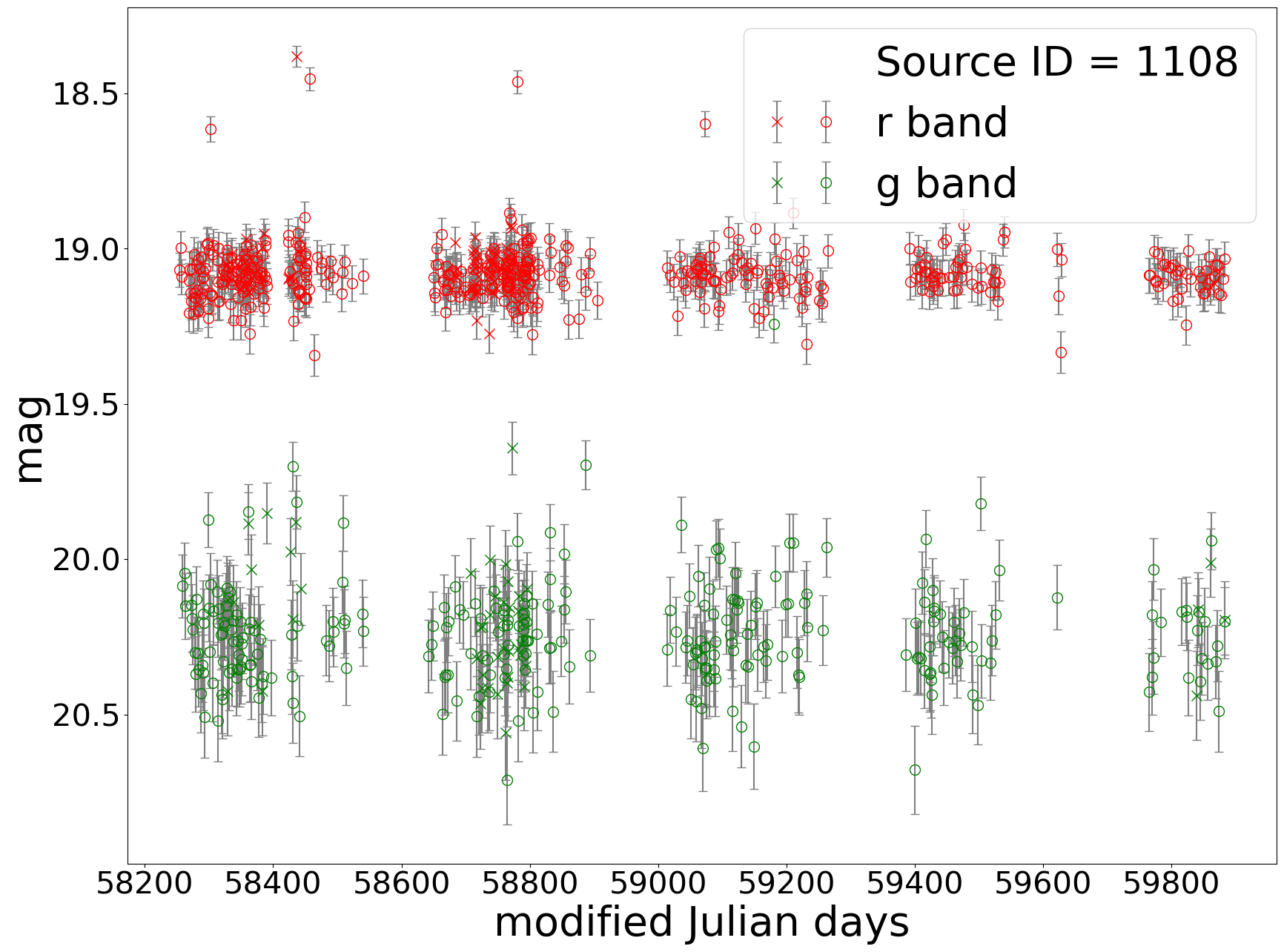}
\caption{Source 1108, a typical flaring super giant. Different markers represents different ZTF labeled OIDs. The survival function $S_k(\chi^2)$ of a constant line model is $8.71\times10^{-14}$ for the $r$ band lightcurve, and $9.07\times10^{-8}$ for the $g$ band lightcurve.}
\label{fig:lc_1108}
\end{figure}

Listed in Table \ref{tab:flaring_sg}, the non-periodic variable sources within IC 10 typically exhibit flaring behavior -- single or a small number of epochs where the source is significantly brighter than preceding or subsequent observations (e.g., Figures  \ref{fig:lc_128} \& \ref{fig:lc_1108}).  As shown in Figure \ref{fig:HRdiagram1}, the location of such sources on the H-R diagram of detected IC 10 sources suggests the majority of these sources are supergiant (SG) stars -- massive stars which have evolved off the main sequence -- a designation supported by the classification of four such objects as Emission line stars in the SIMBAD database (Table \ref{tab:flaring_sg}), one of which (Source 85) is also a candidate Luminous Blue Variable (LBV) star \citep{richardson18}.  Below, we consider different possible origins for the flaring behavior observed from these putative SGs.

\begin{table}
\begin{tabular}{lll}
\hline
\hline
Source ID & Classification & ${\mathcal R}$ \\ \hline
1053 & $\cdots$ & -0.113 \\
\hline
492 & Em*       & 0.026  \\
222 & $\cdots$  & 0.074  \\ 
1108 & $\cdots$ & 0.082  \\ 
247 & $\cdots$  & 0.116  \\
487 & $\cdots$  & 0.118  \\ 
62  & $\cdots$  & 0.122  \\
188 & Em*       & 0.123  \\
184 & Em*       & 0.128  \\ 
293 & $\cdots$  & 0.129  \\ 
375 & $\cdots$  & 0.139  \\
65  & Cl*       & 0.169  \\
331 & $\cdots$  & 0.174  \\
64  & Cl*       & 0.179  \\
883 & $\cdots$  & 0.190  \\ 
\hline
226 & $\cdots$  & 0.269  \\ 
255 & Em*       & 0.273  \\
85  & Em* / LBV & 0.288  \\ 
327 & $\cdots$  & 0.301  \\ 
639 & $\cdots$  & 0.321  \\ 
\hline
269 & $\cdots$  & 0.375  \\
520 & $\cdots$  & 0.391  \\ 
570 & $\cdots$  & 0.408  \\ 
574 & $\cdots$  & 0.426  \\ 
128 & $\cdots$  & 0.439  \\ 
25  & $\cdots$  & 0.445  \\ 
\hline
\hline
\end{tabular}
\caption{Source ID, Classification, and the Pearson correlation coefficient between $r$ and $g$ band magnitudes measured $<1$ day apart, ${\mathcal R}$ of non-periodically variable ZTF sources within the optical extent of IC 10, ordered by increasing values of ${\mathcal R}$.}
\label{tab:flaring_sg}
\end{table}

To determine if the flares observed from these sources were contemporaneously observed in both bands, for each one we calculated the Pearson correlation coefficient $\mathcal{R}$ between the apparent $z$ and $r$ magnitudes measured $<1$~day apart.  As shown in Table \ref{tab:flaring_sg}, this class of 23 sources exhibit a wide range in $\mathcal{R}$, with 12 showing little to no correlation ($|{\mathcal R}| \lesssim 0.2$ between the band, five showing possible but weak correlation ($0.25 \lesssim |{\mathcal R}| \lesssim 0.325$), and six showing a mild correlation ($0.375 \lesssim |{\mathcal R}| \lesssim 0.45$) between the brightness in these two bands.  This diversity suggests possible differences in the physical origin of the flares observed from this collection of objects.

\begin{figure}
 \centering
 \includegraphics[width=0.9\linewidth]{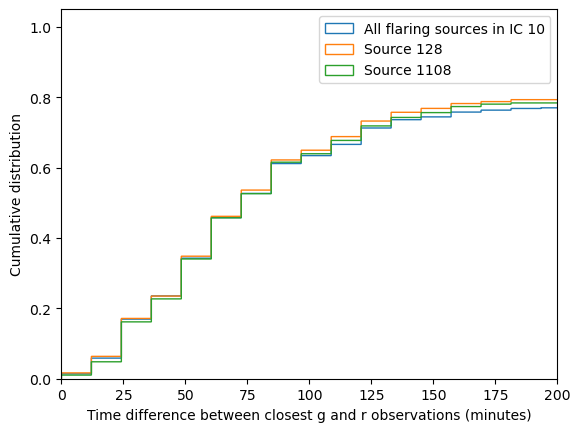}
 \caption{Cumulative distribution of time between the closest $r$ and $g$ band observations for flaring super giants in IC 10.}
 \label{fig:t_diff_r_g}
 \end{figure}

For all of these objects, their $g$ and $r$ band likely originates from thermal processes.  In this case, the change in temperature and/or emitting radius required to vary the intrinsic luminosity of the source would impact the brightness in both bands.  Therefore, the lack of a highly significant correlation between $r$ and $g$ band suggests the duration of the flaring events is less than the time between successive observations in each band.  As shown in Figure \ref{fig:t_diff_r_g}, $\sim80\%$ of $g$ and $r$ observations were $\lesssim3.5~{\rm hours}$ apart -- suggesting an extremely short duration, and corresponding small emitting region, of these flares, at odds with the large size of  supergiant stars.  However, given the significant fraction of SGs found in binary systems -- e.g.,
$\sim 20\%$ of red SGs in the Large Magellanic Cloud are in binaries \citep{neugent20} -- these flares could result from the interaction between the SG and its companion -- likely to be another massive star (e.g., \citealt{neugent20}) or the compact object (i.e., neutron star or stellar mass black hole) produced by the core-collapse of an initially more massive stellar companion. The spatial coincidence between Source 128 and a upernova remnant (as cataloged in SIMBAD; Table \ref{tab:simbadclass}) suggests this scenario is feasible.  A recent study of high-mass X-ray binary IC 10 X-2 found it exhibits short-duration flares similar to those observed from these SGs,  believed the result of clumps of intra-binrary material accreting onto the neutron star in this system \citep{alnaqbi25}.  Future spectroscopic objects would determine if these SGs are in binaries and, if so, the mass of their companions -- testing if the observed flares are indeed the result of an interaction with their companion.

\subsection{Source 193}
\label{sec:193}
\begin{figure}
\centering
\includegraphics[width=0.9\linewidth]{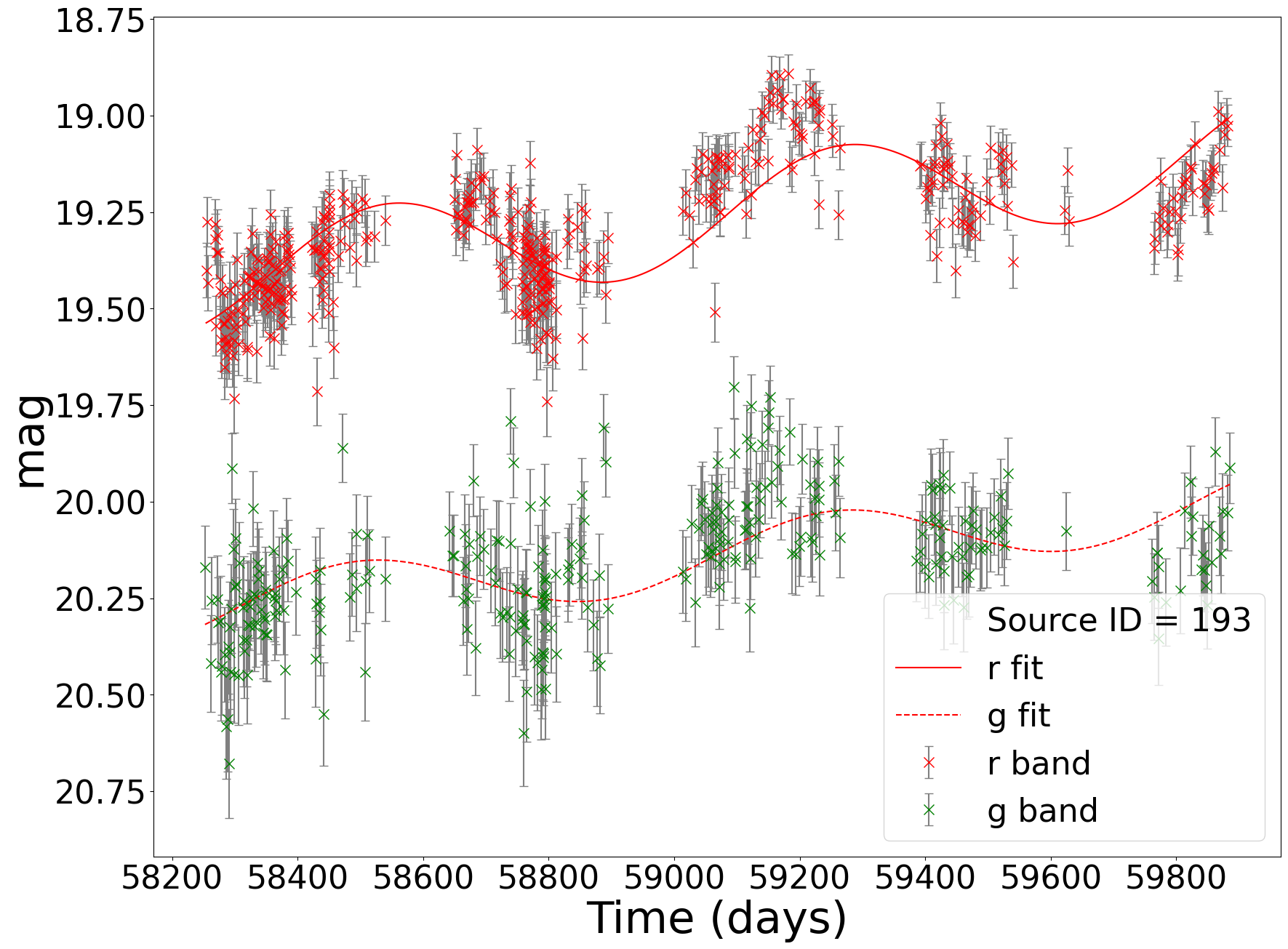}
\caption{The lightcurve of Source 193 for the fit parameters presented in Table \ref{tab:193}.}
\label{fig:lc_193}
\end{figure}

Since Source 193 is classified as a candidate luminous blue variable (LBV) by \citet{richardson18}, its observed emission is of particular interest. While Source 193 is identified as a non-periodic variable within IC 10 and is located within the SG branch on the H-R diagram, as shown in Figure \ref{fig:lc_193}, it's lighcurve is different from the flaring sources outlined above. Instead of a roughly constant magnitude with short duration variations as the sources in \S\ref{sec:ic10-flares}, the brightness of this source is well described by a sinusoidal plus a straight line of decreasing magnitude:
\begin{eqnarray}
\label{eqn:src193lc}
 m & =& A\cdot \sin(\omega\cdot t+\phi_0)+k\cdot t+C
\end{eqnarray}
which the parameters provided in Table \ref{tab:193}. 

The seemingly linear decrease in magnitude might indicate a second periodicity on timescales much longer than the $\sim1600$d duration of the observations presented here. Around one third of long-period variables (LPVs), which are mostly red giants and red supergiants, show long secondary periods (LSPs) that to be one order of magnitude longer than the primary period \citep{Wood_1999, soszyński2022eppursimuoveorigin}. The origin of LSPs in red giants and red supergiants is still uncertain, with possible hypotheses including binary motions, dust formation, rotation, and/or non-radial oscillations\citep{Nicholls_2009,Wood_2004}. However, Source 193 is a candidate LBV, and not a red giants or red supergiants as LPVs with LSPs typically are.  Therefore, its observed long term increase in brightness could perhaps be a precursor to a ``SN impostor'' event, as observed from U2773-OT in UGC 2773 (e.g., \citealt{smith10}).  Further photometric and spectroscopic observations are required to determine the origin of both the observed periodicity and secular increase in brighness observed from this source.

\begin{table*}[]
\footnotesize
\begin{tabular}{llllllll}
Band & $T=\frac{2\pi}{\omega}$ (days) & $A$ & $\omega\times10^{4}$ & $\phi_0$ & $k\times10^{5}$ & $C$ & $\chi^2$ \\ \hline
$r$ & $726.18\pm9.20$   & $0.138\pm0.010$   & $86.52\pm1.10$ & $19.69\pm6.46$  & $-20.90\pm1.36$ & $31.60\pm0.80$ & 2.80 \\ \hline
$g$ & $754.65\pm24.61$  & $0.084\pm0.015$   & $83.26\pm2.71$ & $39.16\pm16.02$ & $-17.19\pm2.06$ & $30.29\pm1.22$ & 1.81 \\ 
\end{tabular}
\caption{\label{tab:193} Best-fit parameters when modeling the apparent magnitude of Source 193 using Equation \ref{eqn:src193lc}, as well as resultant reduced $\chi^2$.}
\end{table*}

\subsection{Periodic super giants}
\label{sec:periodic_sg}
\begin{figure*}
\centering
\includegraphics[width=0.31\linewidth]{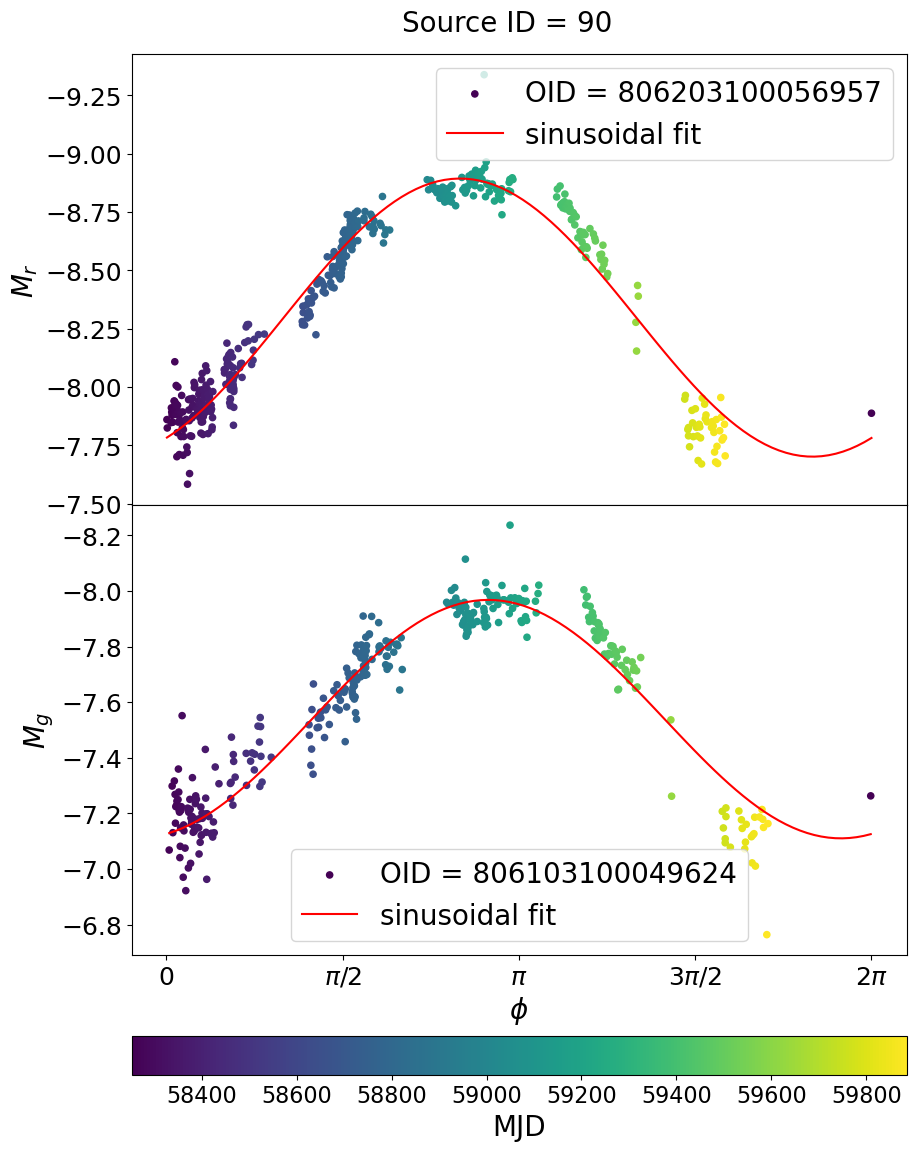}
\includegraphics[width=0.31\linewidth]{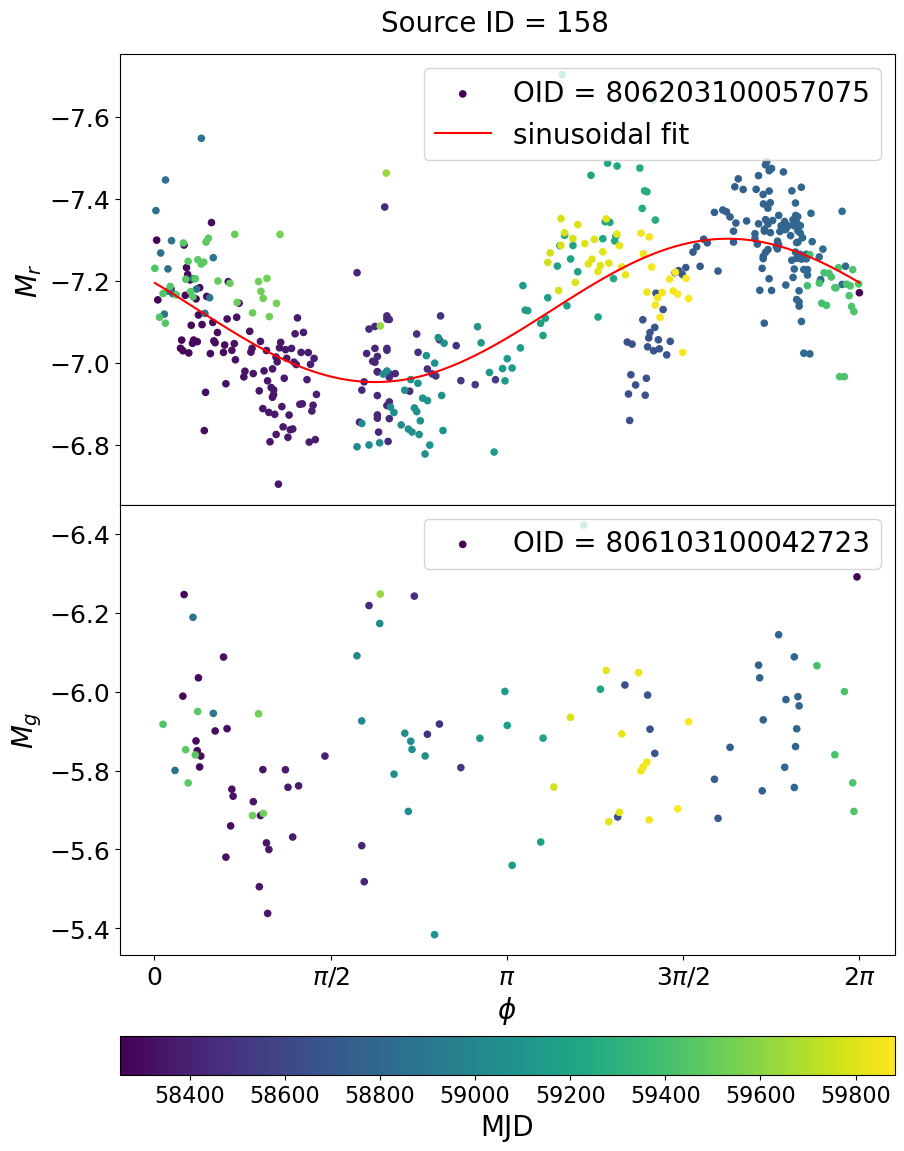}
\includegraphics[width=0.31\linewidth]{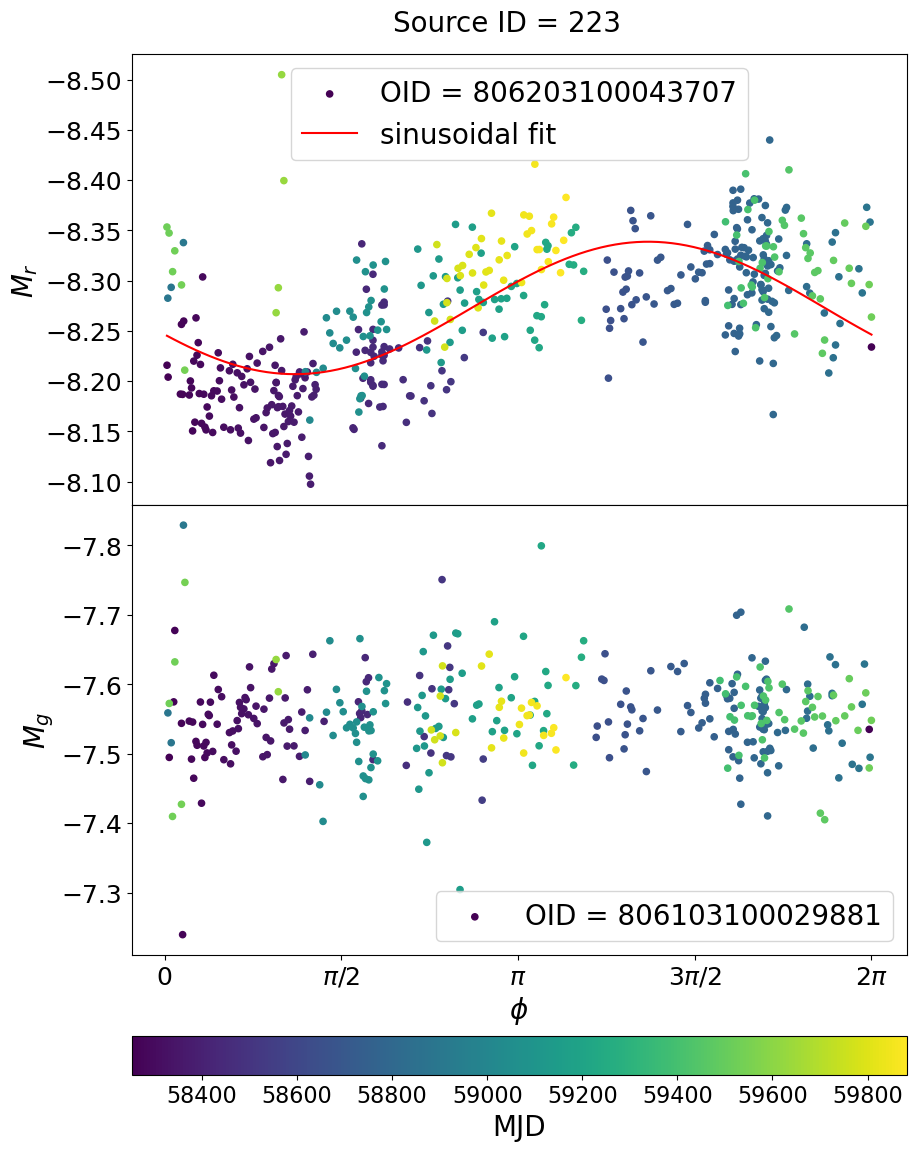}
\includegraphics[width=0.31\linewidth]{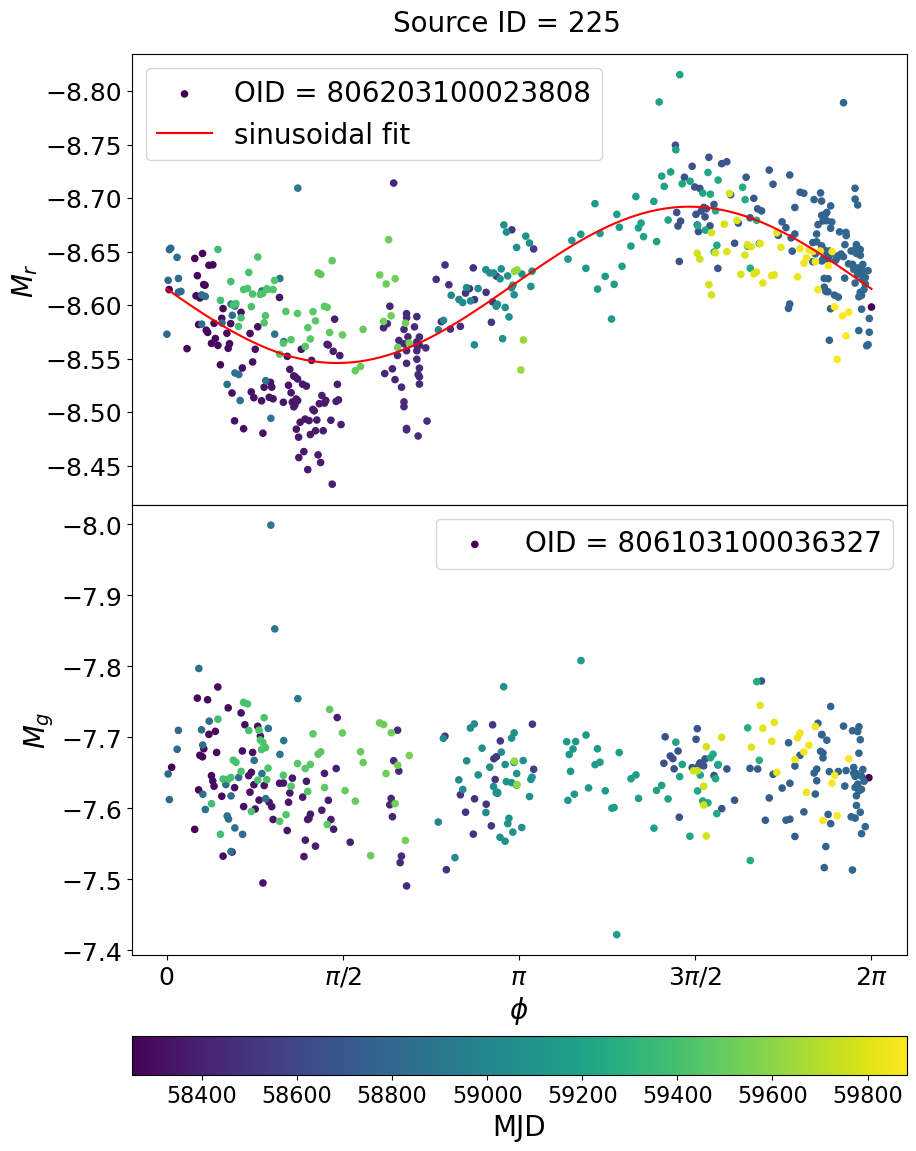}
\includegraphics[width=0.31\linewidth]{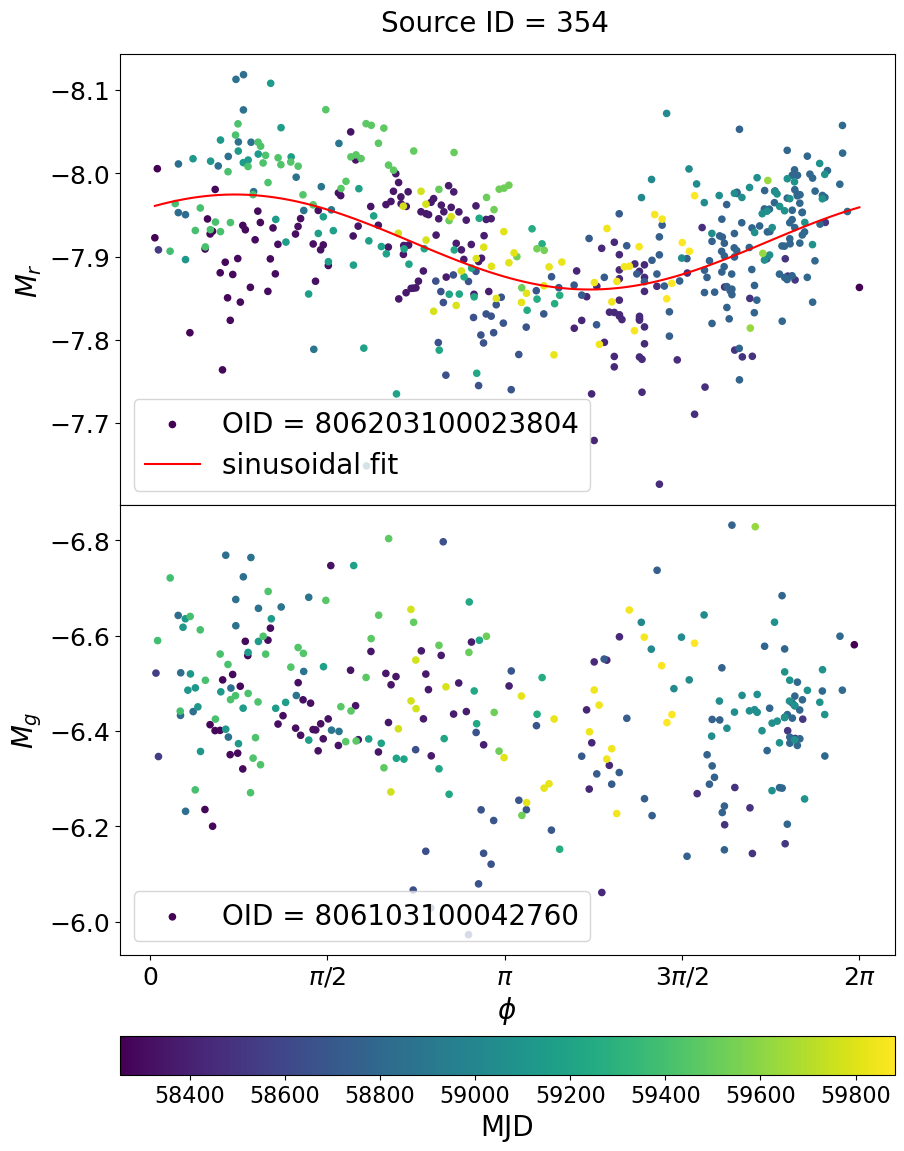}
\caption{Folded lightcurves of periodic super giants with a sinusoidal fit for bands with statistically significant periodicity. The lightcurves are folded by the Lomb-Scargle period if periodic, or by the Lomb-Scargle period of the other band if non-periodic. The color denotes the MJD of the observation.}
\label{fig:periodic_sg}
\end{figure*}

In addition to the sources discussed above, there are also six periodically varying sources among the SG branch within IC 10 (see Table \ref{tab:periodic_sg}).  One of these, Source 77, is the previously identified High Mass X-ray Binary IC10 X-2, whose variability is analyzed in detail by \cite{alnaqbi25}.  In this section, we discuss the nature of the other five sources, which have periods ranging from $\sim275 - \sim2000$d (Table \ref{tab:periodic_sg}), and whose folded light curves (Fig.~\ref{fig:periodic_sg}) are well fit by a sinusoidal with a peak $r$ band absolute magnitude $M_{\rm peak,r} \sim -8 - -9$ (Table \ref{tab:periodic_sg}).

\begin{figure}
\centering
\includegraphics[width=0.9\linewidth]{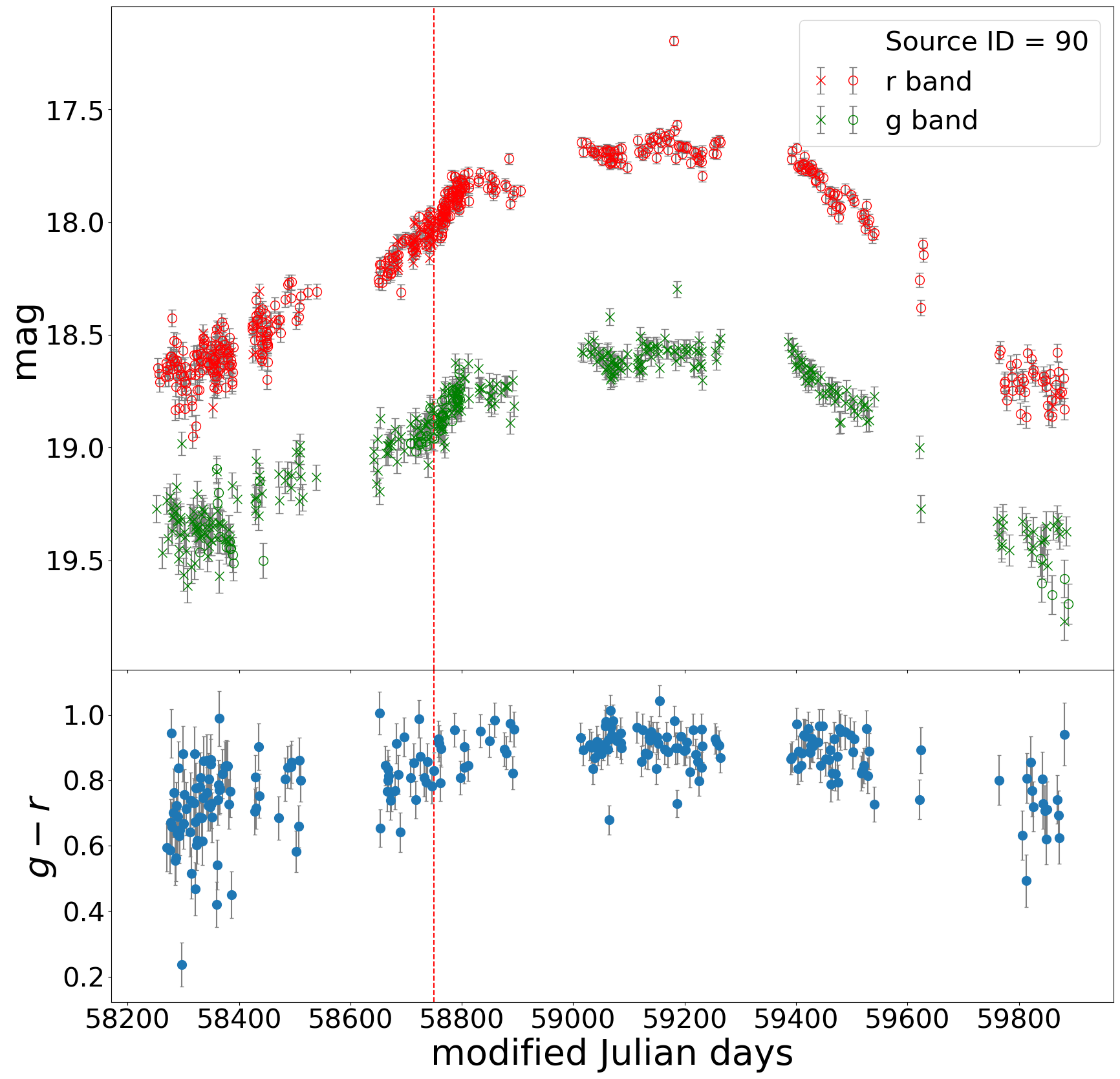}
\caption{Magnitude and color (g-r) of Source 90, with he red vertical line indicates MJD 58750. Different marker represents different ZTF labeled OIDs.
}
\label{fig:lc_color_90}
\end{figure}

\begin{figure}
\centering
\includegraphics[width=0.9\linewidth]{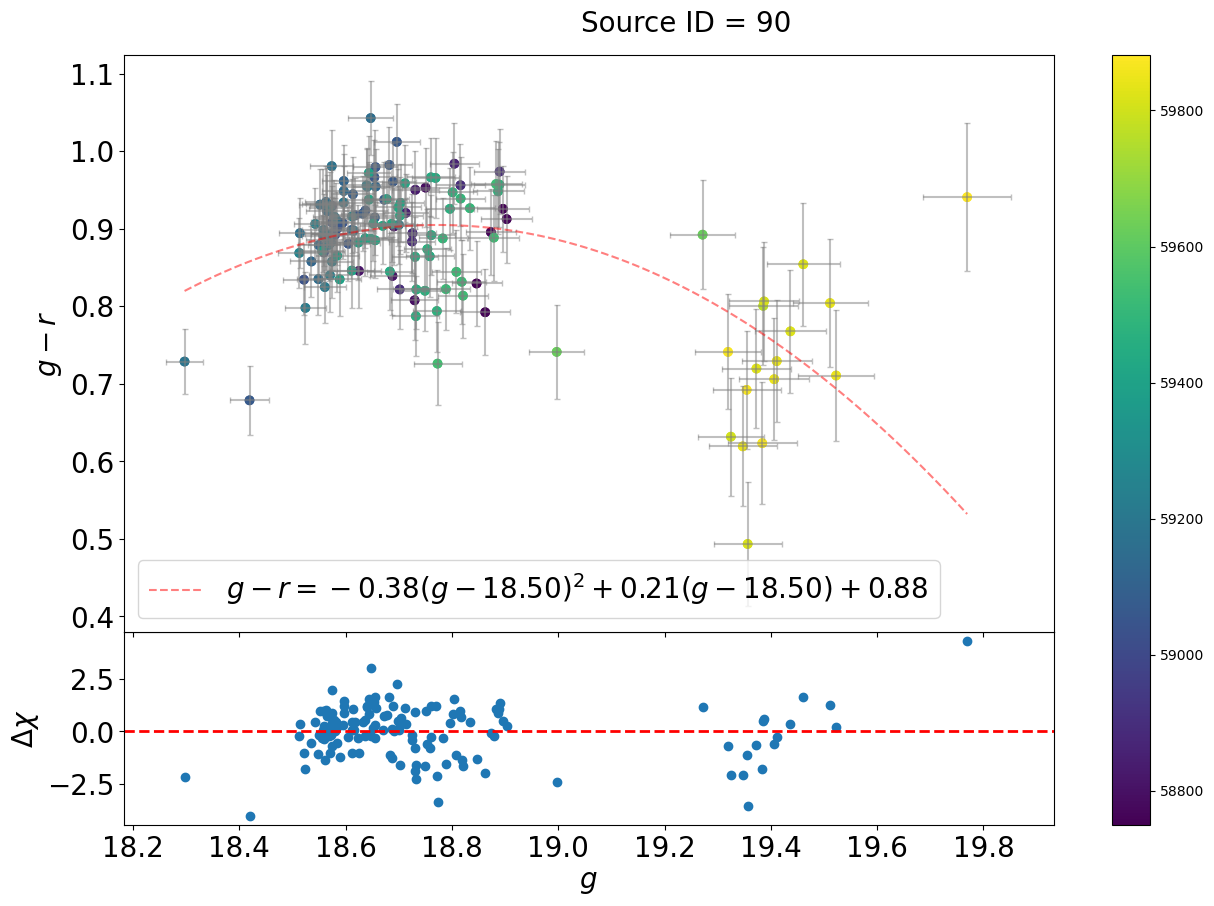}
\caption{$g-r$ color with $g$ band magnitude for source 90 observed after MDJ 58750, fitted by a quadratic polynomial. The lower panel shows the $\Delta \chi^2$ of the fit.}
\label{fig:g_g-r_90}
\end{figure}

Both the peak absolute magnitude and period of these sources is similar to that of S Doradus-type outbursts from LBVs, which typically have $M_\textrm{peak} \sim -8 - -12$ and last $\sim60 - 6000$ days \citep{Smith_2011, Humphreys_1994, Groh_2009}.
A characteristic of such outbursts is that the LBV becomes redder when brighter, since the star cools as it expands during the outburst \citep{Joshi_2019}.
As shown in Figure \ref{fig:lc_color_90}, the source did appear to get redder (an increase in $g-r$) as it got brighter (decreasing magnitude), this trend became more pronounced after MJD 58750 (Figure \ref{fig:g_g-r_90}), which marks the beginning of a more rapid increase in brightness (Figure \ref{fig:lc_color_90}).  
Following previous analyses of similar objects (e.g., \citealt{Sholukhova_2011, Solovyeva_2019, Joshi_2019}), we fit its observed color as a quadratic function of $g$ magnitude during this period:
\begin{eqnarray}
    g-r & = & A(g-g_0)^2 + B(g-g_0) + C,
\end{eqnarray}
where $g_0 \equiv 18.5$, $A=-0.38$, $B=0.21$, and $C=0.88$.  The results of this fit are shown in Fig.~\ref{fig:g_g-r_90}.  
Moreover, the sinusoidal fit (see Table \ref{tab:periodic_sg}) to the light curves suggests the brightness changes by $A \sim 1$~magnitude, comparable to the typical brightness variations during an S Doradus episode of 1–2 mag \citep{Lovekin_2014}.
The $M_\textrm{peak}$, duration, color-magnitude evolution, and change in brightness all suggests that Source 90 is likely a previously unidentified LBV in IC 10.

As mentioned above, for the other four sources (158, 223, 225, 354) the $g$ band lightcurve did not pass our survival function test for a constant line model, thus their $g$ magnitude observations are not sensitive enough to detect any periodic variations. 
However, their detected variations in magnitude ($\sim 0.1$) at $r$ band are smaller than typical S Doradus variations. Although some LBVs do show ``microvariability'', which is a variation of typically a few tenths of a magnitude, such LBVs appear to have stochastic and irregular periods from weeks to months \citep{Abolmasov_2011,Lovekin_2014}. The long and relatively stable periods of these four sources (see Fig.~\ref{fig:periodic_sg}) suggests they are not LBVs with microvariability. While these four sources are LPVs, future observations are needed to determine their nature.

\begin{table*}[]
\footnotesize
\begin{tabular}{llllllllll}
ID & Band & Period (days) & $A$       & $\phi_0$   & $C$ & $\chi^2$ & $t_0$ (MJD) & $M_{\textrm{peak}}$ & SIMBAD\\ \hline
90  & $r$ & 2053.73 & $0.596\pm0.007$ & $2.095\pm0.010$   & $-8.298\pm0.005$ & 8.34     & 58254.49  & -8.894 & Em*   \\
    & $g$ & 1911.34 & $0.429\pm0.007$ & $1.843\pm0.017$    & $-7.539\pm0.005$ & 3.11     & 58252.47 & -7.968 &       \\ \hline
158 & $r$ & 590.25  & $0.175\pm0.008$ & $-0.398\pm0.050$   & $-7.128\pm0.006$ & 3.59     & 58254.49  & -7.303 & Em* \\ 
    & $g$ & - & - & - & - & - & - & - & \\ \hline
223 & $r$ & 633.92  & $0.066\pm0.003$ & $0.414\pm0.050$    & $-8.273\pm0.002$ & 2.41     & 58254.49  & -8.339  & Cl* \\ 
    & $g$ & - & - & - & - & - & - & - & \\ \hline
225 & $r$ & 548.48  & $0.073\pm0.003$ & $0.051\pm0.038$    & $-8.619\pm0.002$ & 2.21     & 58254.49  & -8.692  & Star \\ 
    & $g$ & - & - & - & - & - & - & - & \\ \hline
354 & $r$ & 282.22  & $0.057\pm0.005$ & $-2.323\pm0.070$   & $-7.917\pm0.003$ & 2.45     & 58254.49  & -7.974  & - \\
    & $g$ & - & - & - & - & - & - & - & \\
\end{tabular}
\caption{\label{tab:periodic_sg}Periodic super giants, sinusoidal fit $M=A\cdot \sin(\omega\cdot t+\phi_0)+C$, with $\omega=2\pi/\text{Period}$. The Lomb-Scargle period, fitting parameters, $\chi^2$ of the fit, the starting date of observation $t_0$ are presented in the table. $M_{\textrm{peak}}$ is defined by $M_{\textrm{peak}}=C-A$. Table cells are left empty if the source is non-periodic in the particular band, or no counterpart found in SIMBAD. Information about source 77,``IC10 X-2'', is covered by \cite{alnaqbi25}.}
\end{table*}

\subsection{Mira Super red sources}
\label{sec:superred}
\begin{figure*}
\centering
\includegraphics[width=0.31\linewidth]{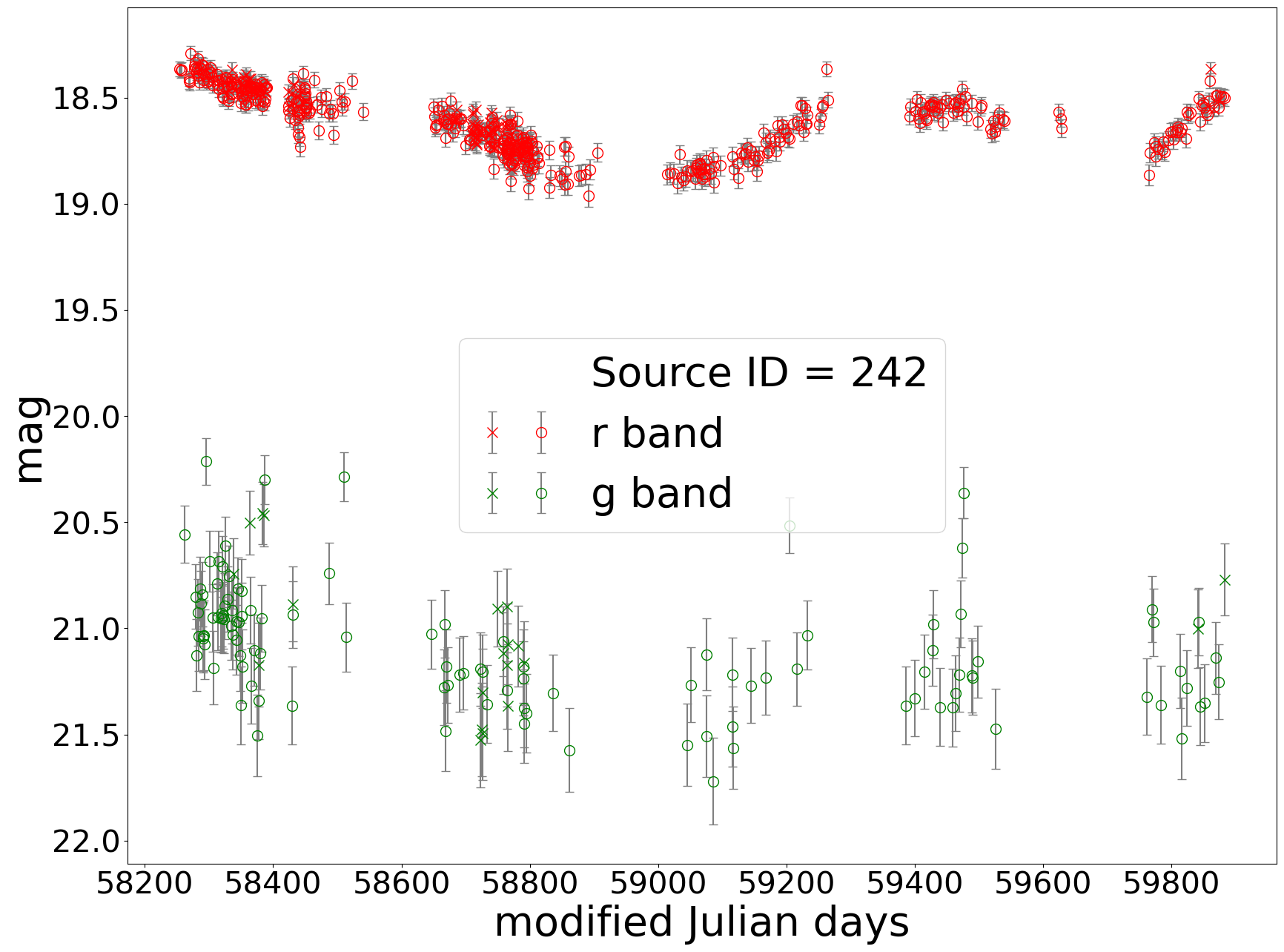}
\includegraphics[width=0.31\linewidth]{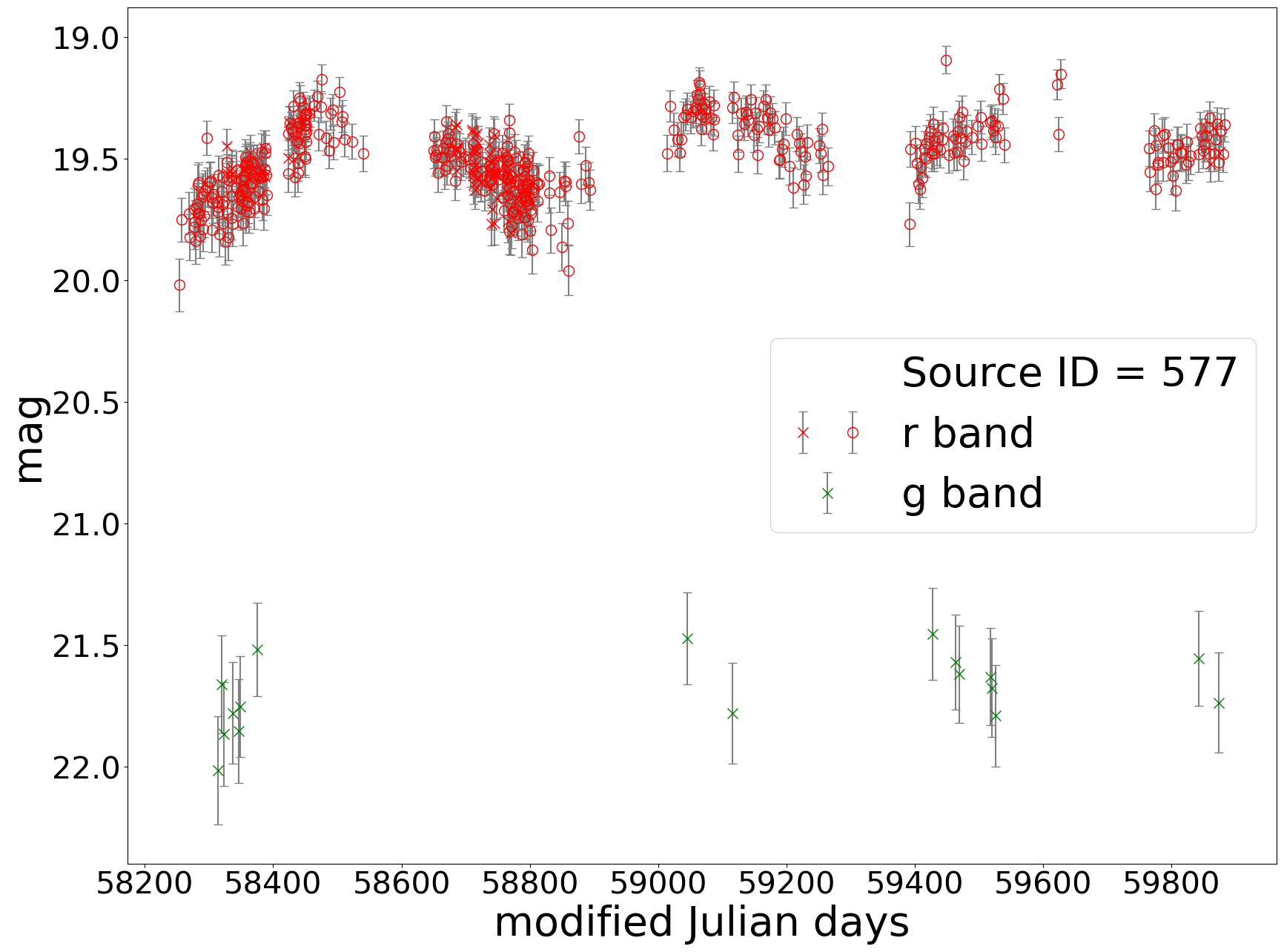}
\includegraphics[width=0.31\linewidth]{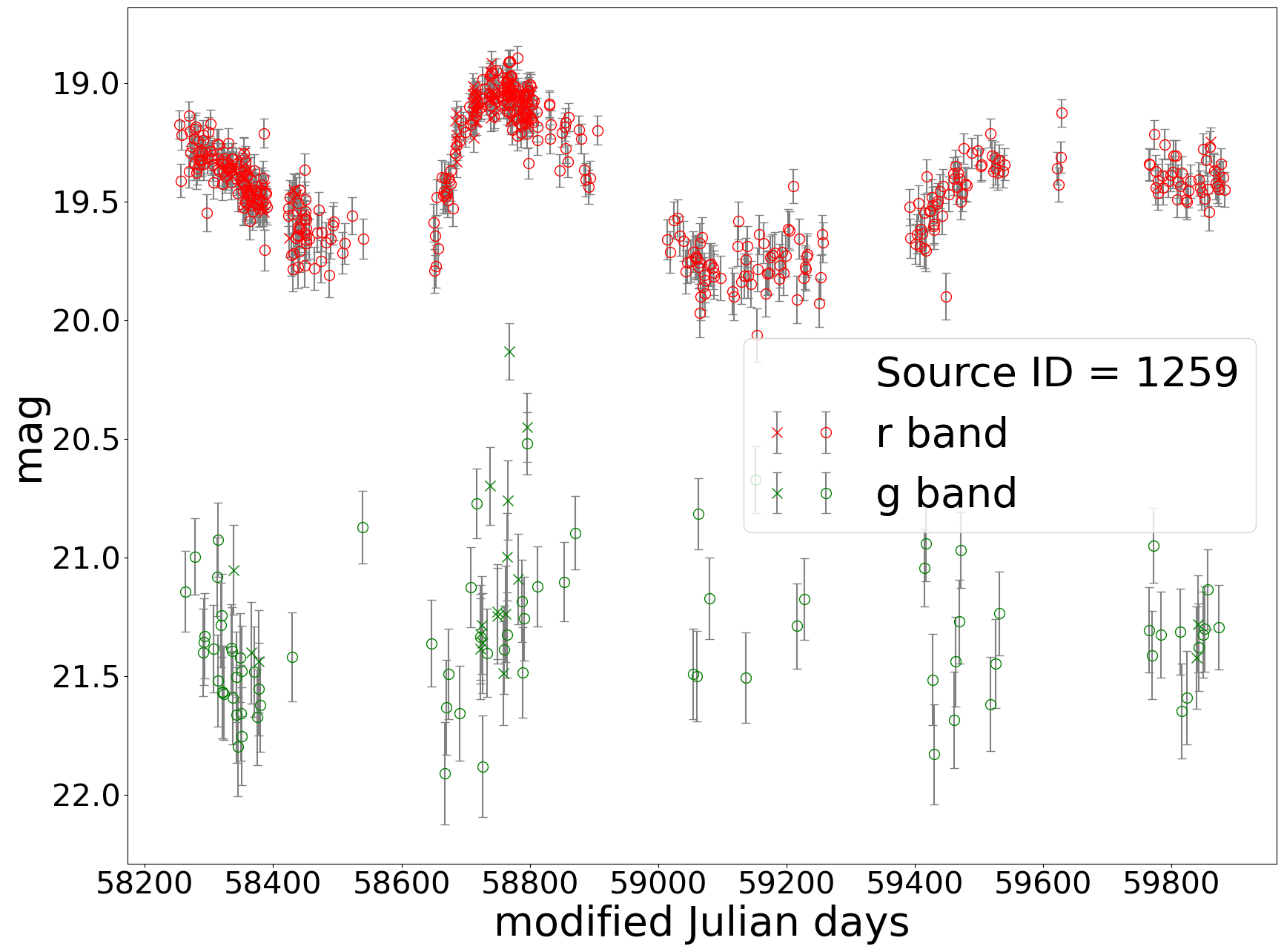}
\caption{Source 242, 557, and 1259, three super red source that are detected in both $r$ and $g$ band. These sources are periodic in $r$ band with a period of 1608, 514, and 595 days respectively. Different markers represents different ZTF labeled OIDs under the same Source ID.}
\label{fig:super_red_miras}
\end{figure*}

As shown in both the color-magnitude diagram (Figure \ref{fig:cmd}) and the H-R diagram (Figure \ref{fig:HRdiagram1}) of the ZTF objects in this field, there are three IC 10 sources detected in both $r$ and $g$ band with $\langle g-r \rangle\ >2$, hereafter referred to as ``super red sources,'' whose demographics are given in Table \ref{tab:superred}.  The emission from all three sources are observed to periodically variable in $r$ band, as shown in their light curves (Figure \ref{fig:super_red_miras}). 
Both their long periods ($P = $1608, 514, and 595 days) and their location on H-R diagram both suggest they are likely Mira stars \citep{bernhard21}.  Mira stars are of particular interest since the observed correlation between the period and luminosity of these objects allows one to determine the distance to these sources -- e.g., \citealt{Ngeow_2023}) determined a period-luminosity relationship for Mira stars are at maximum light in ZTF $r$ and $g$ band, though these relations are only applicable to Miras with periods shorter ($<300$ days) than those we detected in IC 10. \cite{Ngeow_2023} also suggest that long-period (2.48 $<$ log P $\lesssim$ 2.8, or 300 days $<$ P $\lesssim$ 630 days, applicable to the later two sources of the three) O-rich Miras seems to have a near constant or mildly period-dependent r-band absolute magnitude of around $-5$ mag  at maximum light. Our $r$ and $g$ super red sources have an $r$ band apparent magnitude at maximum light of 18.5, 19.3 and 19.2 respectively. If the later two are O-rich Miras and therefore applicable to the constant PL relation, the relation gives distance modulus of 24.3 and 24.2, which agree well with the IC 10 distance modulus of $\sim 24$ found in literature \citep{Sakai_1999,2005ARBl...20...85O,Kim_2009}.  Future spectroscopic studies are needed to both confirm the identification of these objects as Mira stars and, if so, determine their sub-type.

\section{Foreground sources}
\label{sec:othersources}

Objects not associated with IC 10 are likely to be foreground, i.e. Milky Way,  objects or more distant galaxies.  In this section, we discuss the properties of subsets of these sources. 

\subsection{H-R diagram}
\begin{figure}
\centering
\includegraphics[width=0.9\linewidth]{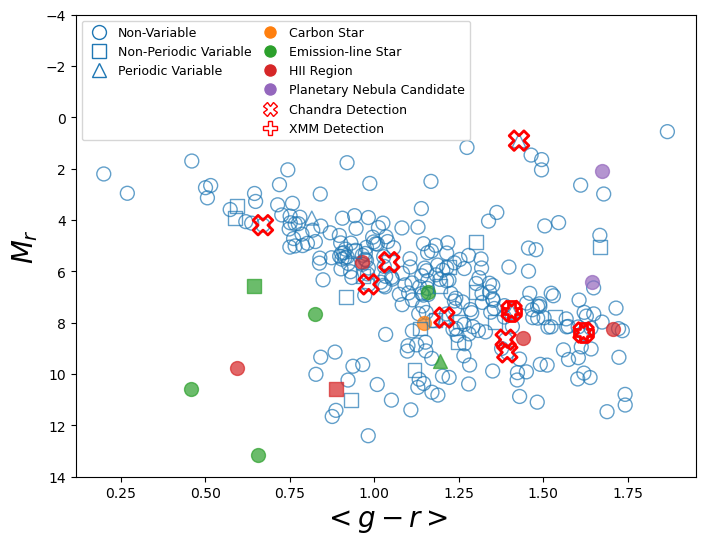}
\caption{H-R diagram for foreground sources. The marker shape shows the variability, the color code denotes SIMBAD identification, and the red symbols indicate X-ray emission as detected by {\it Chandra} (``X'') or {\it XMM} (``+'').}
\label{fig:HRdiagram2}
\end{figure}

For ZTF sources whose {\it GAIA} counterparts have measured parallaxes, we can calculate their absolute magnitude using the the inferred distance -- which places them within the MW.  The resultant HR diagram of these sources detected in both bands, indicates the foreground sources identified in this field are -- perhaps not unsurprisingly -- primarily main sequence stars or white dwarfs (Figure \ref{fig:HRdiagram2}). 

\subsection{Super red sources detected only in $r$ band}
\label{sec:superred2}
\begin{figure*}
\centering
\includegraphics[width=0.31\linewidth]{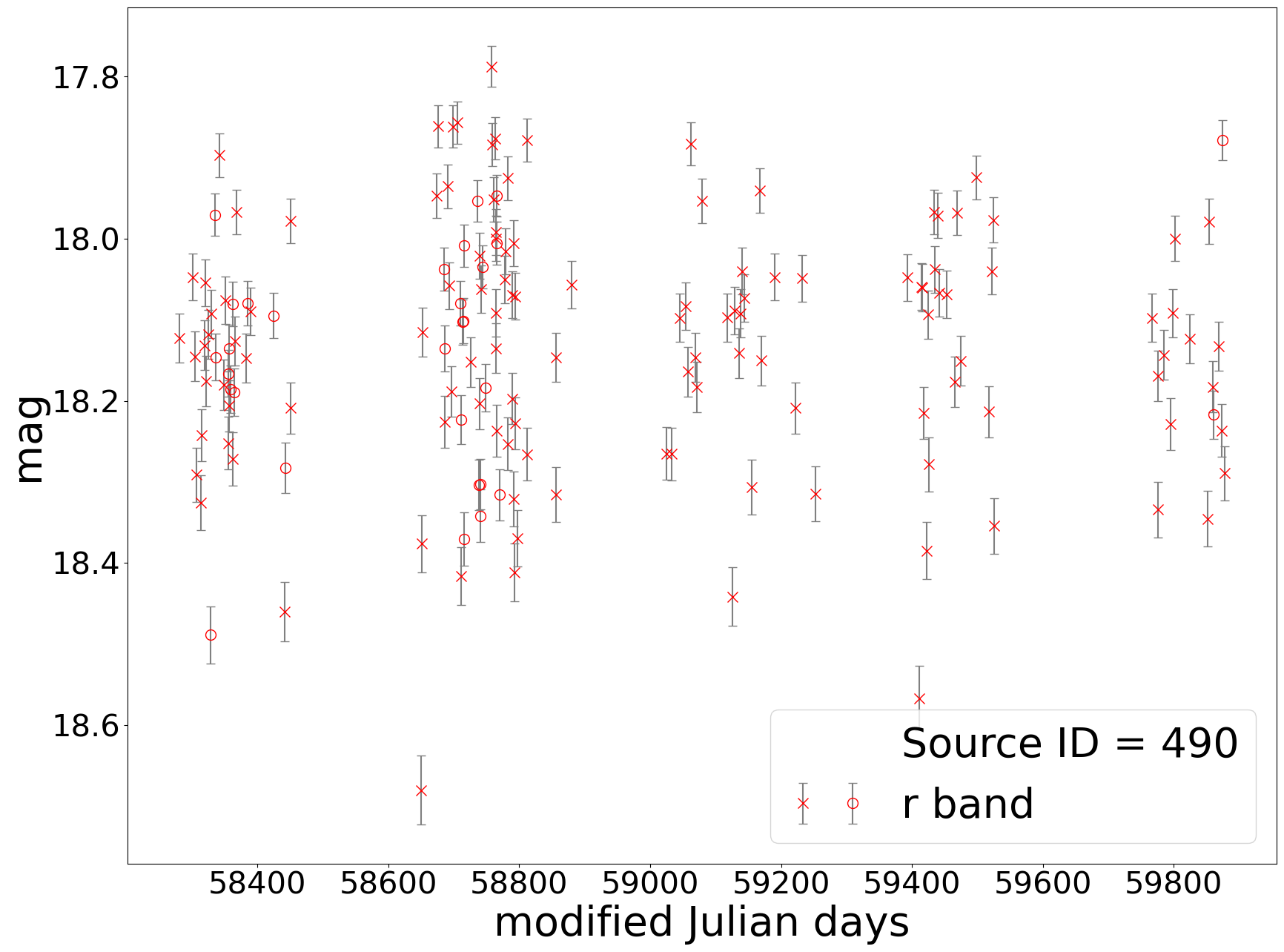}
\includegraphics[width=0.31\linewidth]{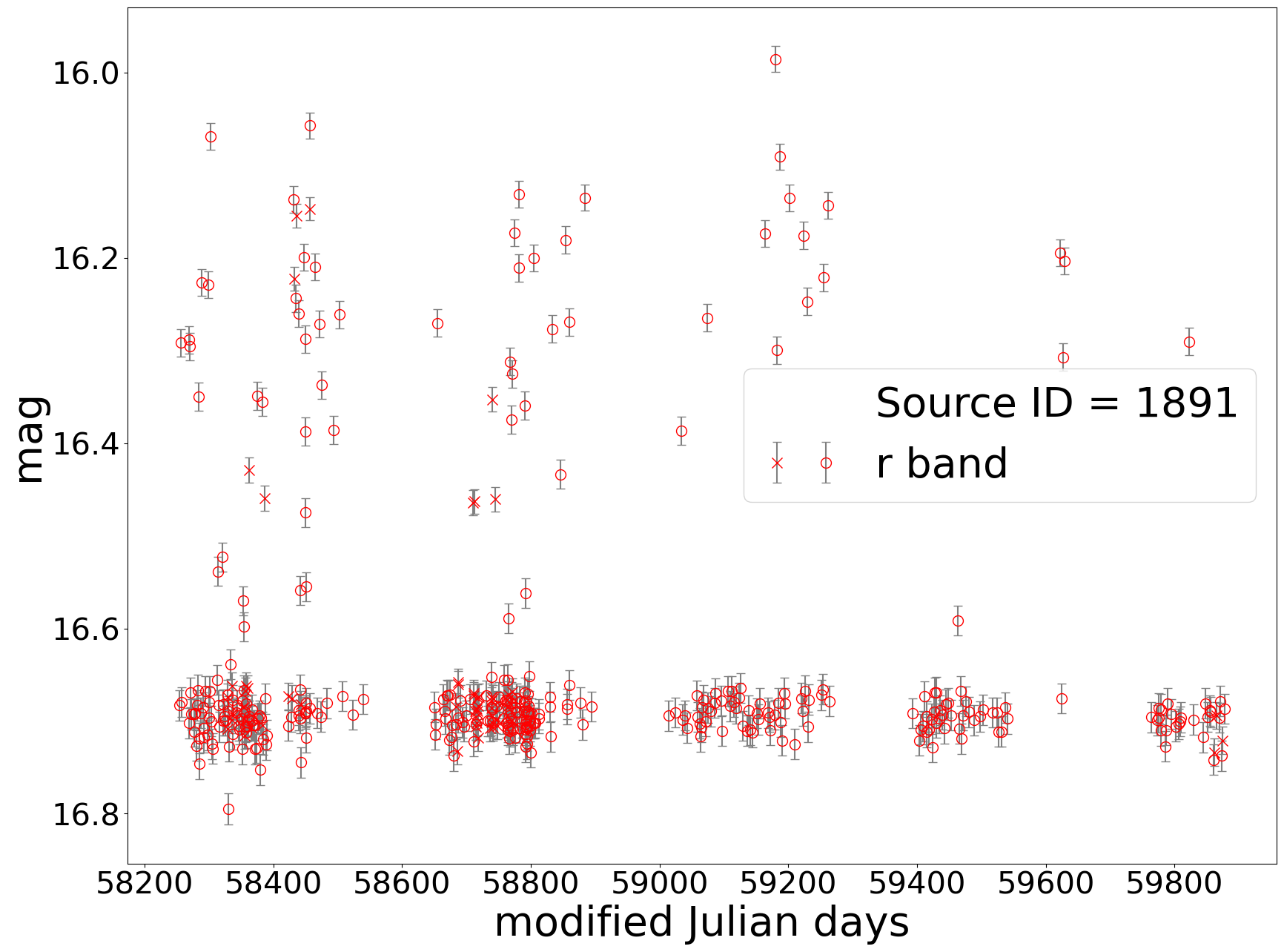}
\includegraphics[width=0.31\linewidth]{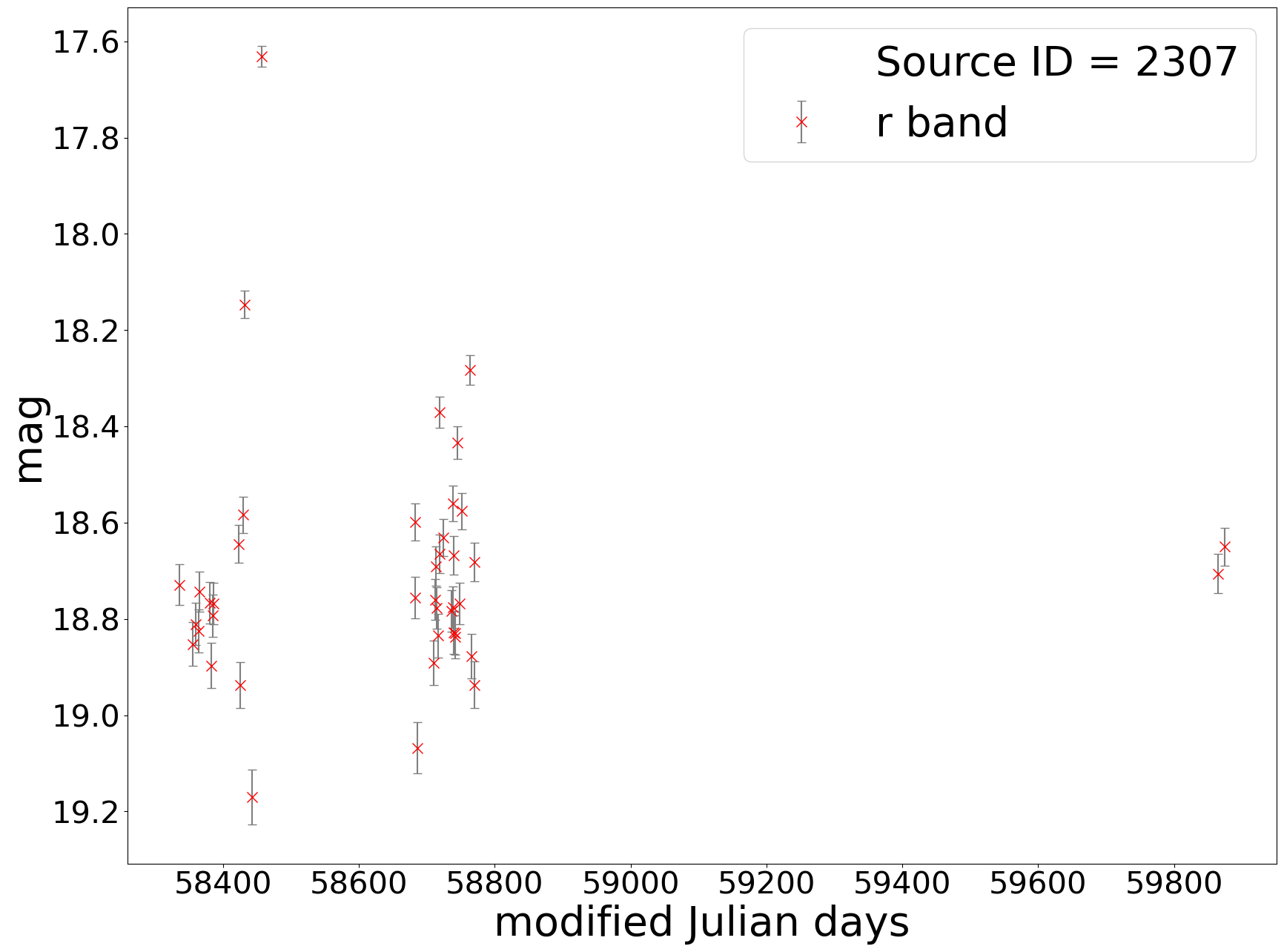}
\includegraphics[width=0.31\linewidth]{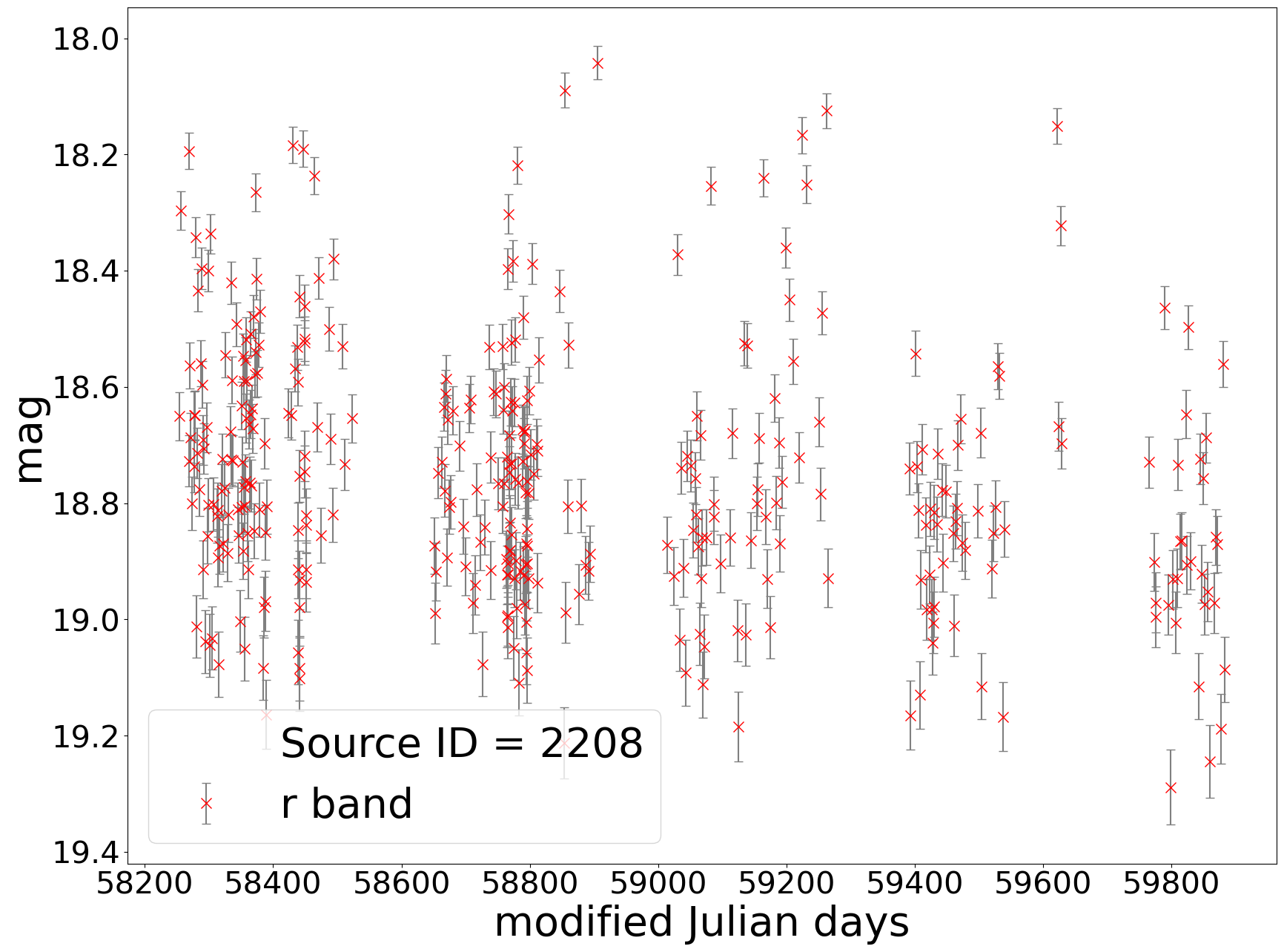}
\includegraphics[width=0.31\linewidth]{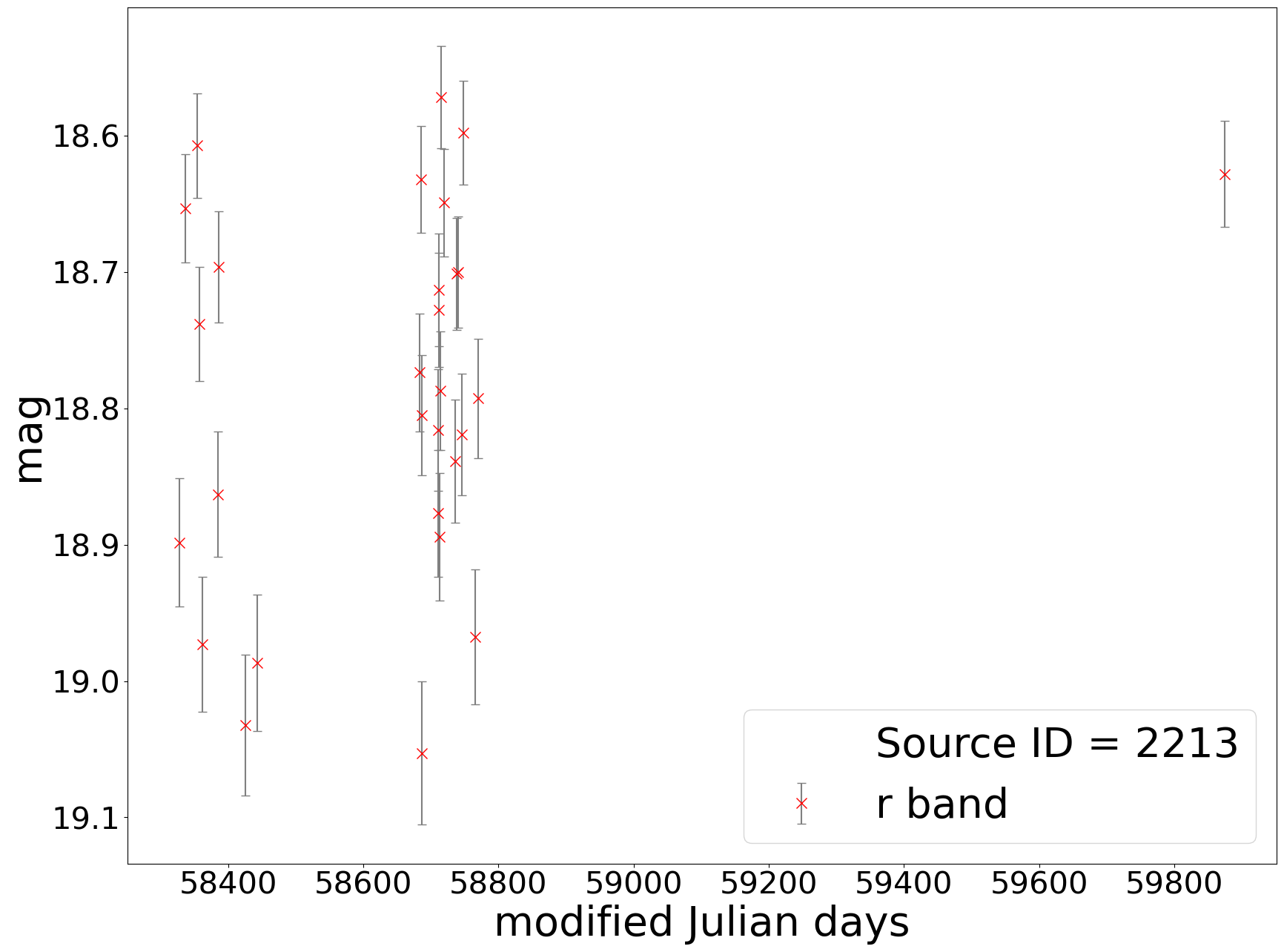}
\includegraphics[width=0.31\linewidth]{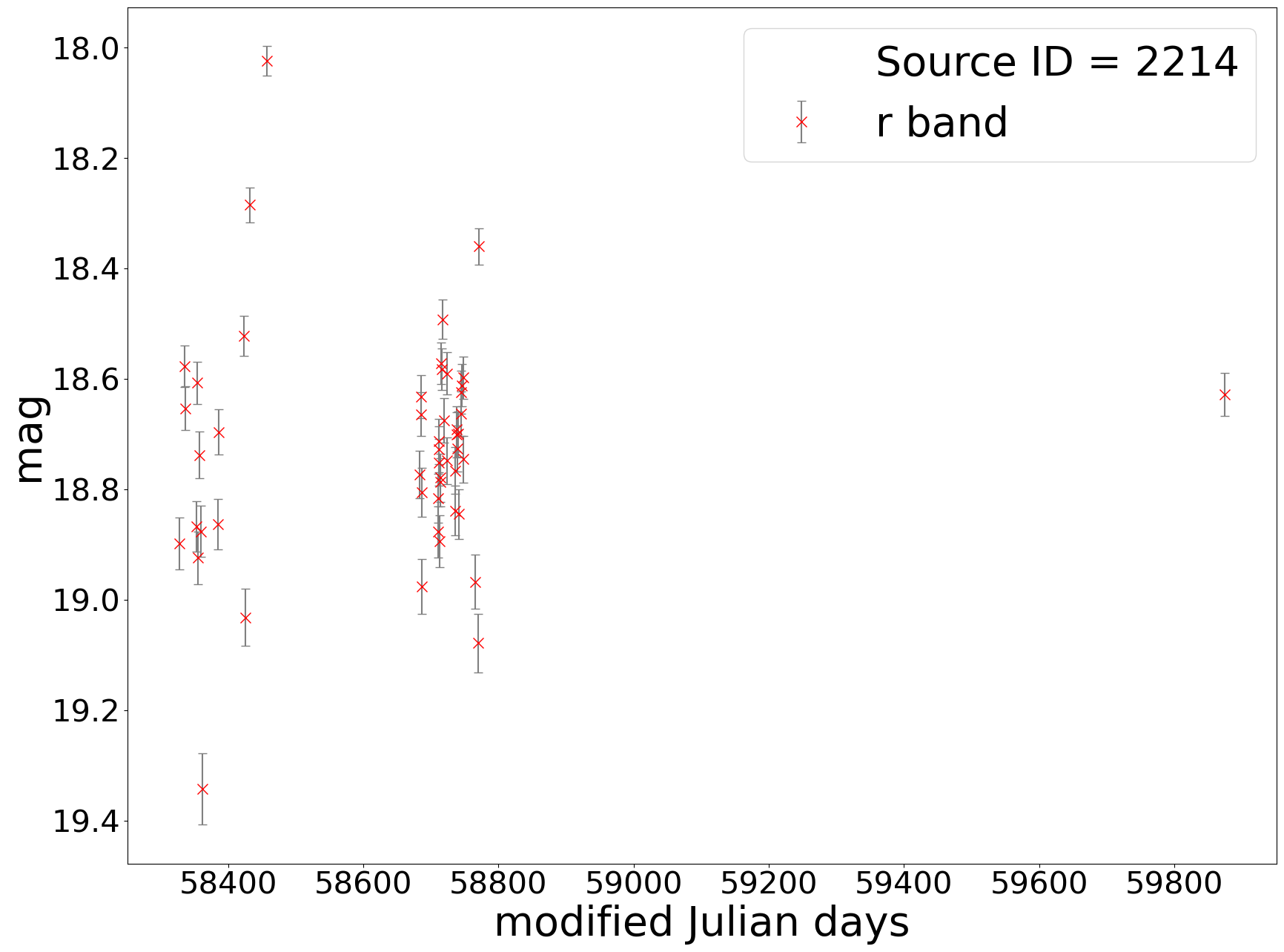}
\includegraphics[width=0.31\linewidth]{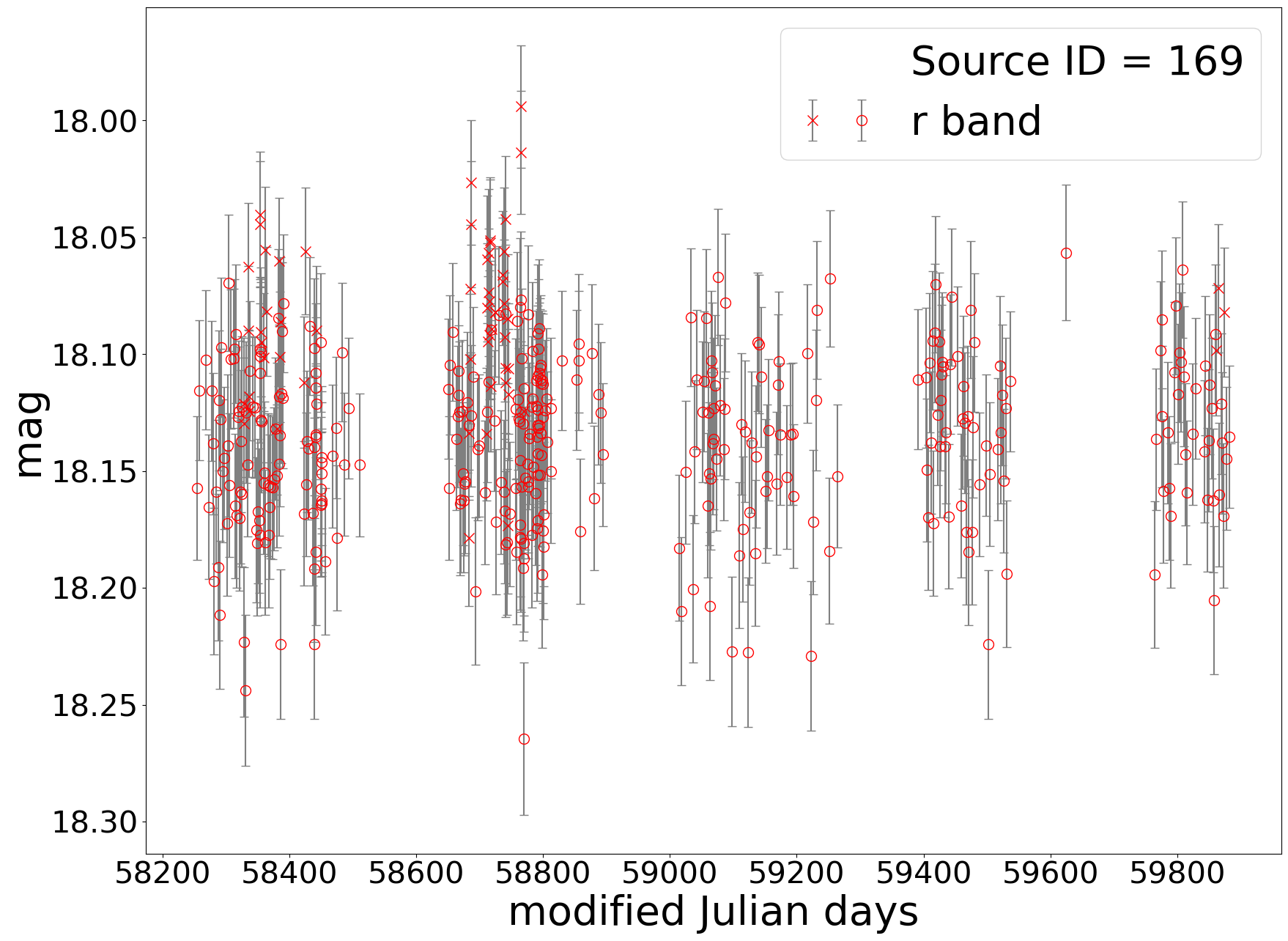}
\includegraphics[width=0.31\linewidth]{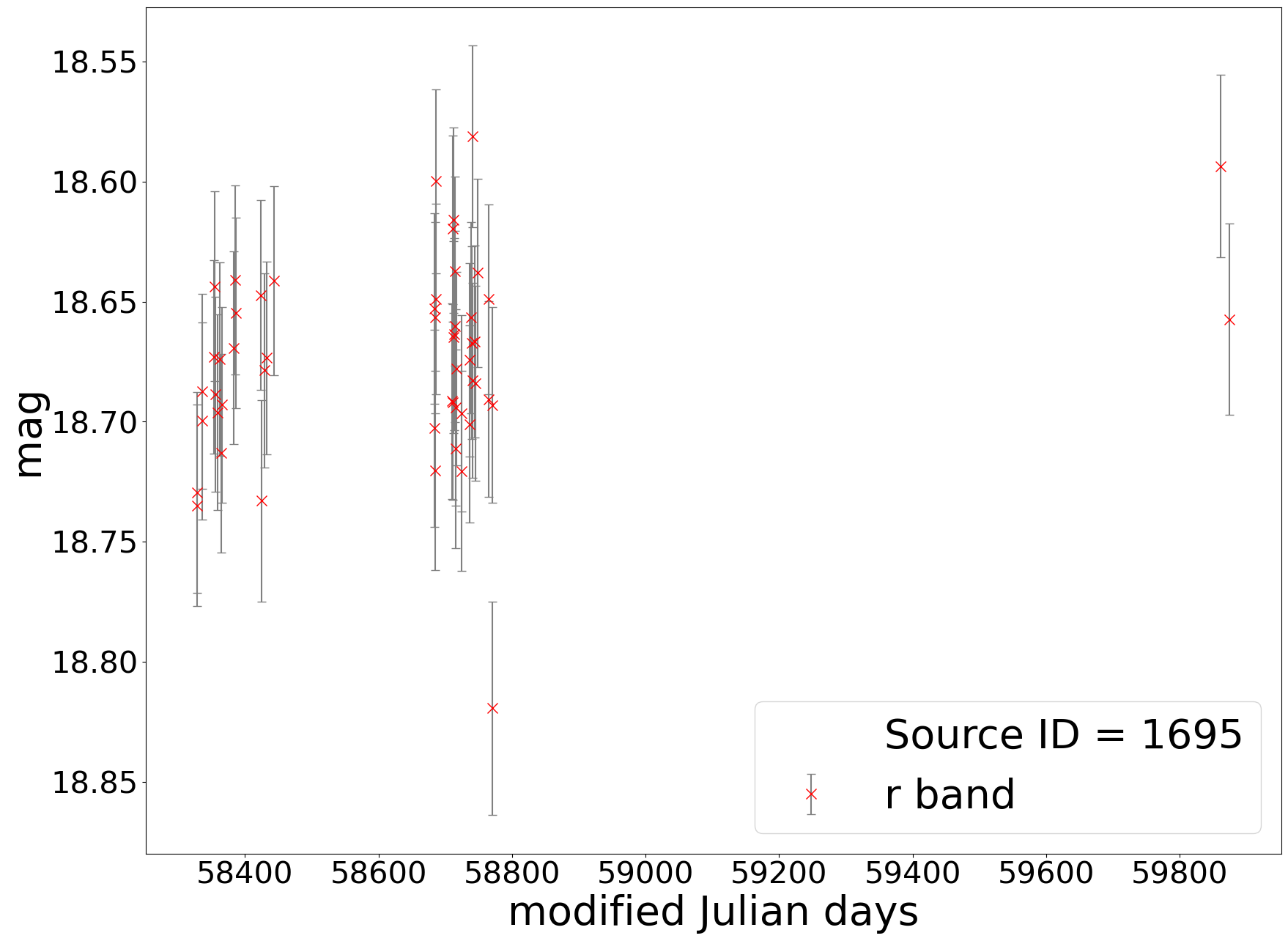}
\caption{\textbf{$1^{\text{st}}$ row}: Three super red non-periodic variable sources detected only in $r$ band and have {\it Gaia} matches. The average absolute magnitude these sources (Source 490, 1891, and 2307) are 7.58, 4.26, and 10.20 mag.
\textbf{$2^{\text{nd}}$ row}: Three super red non-periodic variable sources detected only in $r$ band but do {\it do not} have {\it Gaia} matches.
\textbf{$3^{\text{rd}}$ row}: Two super red non-variable sources detected only in $r$ band and have {\it Gaia} matches. The average absolute magnitude these sources (Source 169 and 1695) are 7.07 and 7.55 mag.
Different markers represents different ZTF labeled OIDs under same Source ID.
}
\label{fig:super_red_foreground}
\end{figure*}

Among these foreground objects are additional ``super red sources" $\langle g-r \rangle\ >2 $, identified by a non-detection in $g$ band -- suggesting their $g$ magnitude is $>20.8$, the limiting magnitude in this band \citep{Bellm_2019} -- but have an average r-magnitude of $\langle r \rangle\ >18.8$.  The properties of these eight sources are given in Table \ref{tab:superred} and, as shown in their light curves (Figure \ref{fig:super_red_foreground}), and all but two of them are non-periodically variable.  

Five of these super red objects are coincident with {\it Gaia} sources that have a measured parallax, allowing us to determine their distance and absolute magnitudes (Table \ref{tab:superred2}), allowing us to constrain their location on the H-R diagram (Figure \ref{fig:HRdiagram2} 
Based on this analysis, Source 1891 ($\langle M_r \rangle =4.26$) is likely to be a giant or subgiant star, Sources 169, 490, and 1695 (all with $\langle M_r \rangle \sim 7.5$ are probably main sequence stars, while Source 2307 $(\langle M_r  \rangle \approx 10.20)$ is likely a red dwarf.  The short duration increases in brightness observed from Source 1891 is similar that observed from other giant stars (e.g., \citealt{olah22}), which the $\sim0.1$~mag spread in the apparent magnitudes of candidates M-dwarfs Sources 490 and 2307 is larger than typically observed from local examples of these objects (e.g., \citealt{hosey15}).

\begin{table*}
\resizebox{\linewidth}{!}{
\begin{tabular}{llll}
                        & Mira Super red sources in both $r$ and $g$ & Super red sources detected only in $r$ & Super red sources in total  \\ \hline
Total                   & 3                                     & 8                                      & 11                          \\ \hline
Periodic Variable       & 3                                     & 0                                      & 3                           \\ \hline
Non-Periodic Variable   & 0                                     & 6                                      & 6                           \\ \hline
Non-Variable            & 0                                     & 2                                      & 2                           \\ \hline
Foreground/Gaia matches & 0/3                                   & 5/5                                    & 5/8                         \\
\end{tabular}
}
\caption{\label{tab:superred}Demographics for super red sources.}
\end{table*}

\begin{table*}[]
\begin{tabular}{lllllll}
SourceID & RA        & DEC          & Var & $<m_r>$ & Distance (pc)      & $<M_r>$     \\ \hline
2208     & 5.0780839 & 59.3143299   & V   & 18.74   & -                  & -           \\ \hline
2213     & 5.0775472 & 59.3150577   & V   & 18.79   & -                  & -           \\ \hline
2214     & 5.0787745 & 59.3148923   & V   & 18.73   & -                  & -           \\ \hline
169      & 5.0788615 & 59.3172927   & N   & 18.14   & $1637.47\pm311.83$ & $7.07\pm0.41$\\ \hline
490      & 5.1145497 & 59.2946405   & V   & 18.12   & $1286.34\pm467.11$ & $7.58\pm0.79$\\ \hline
2307     & 5.118355  & 59.2939669   & V   & 18.70   & $500.33\pm220.61$  & $10.20\pm0.96$\\ \hline
1695     & 5.173176  & 59.3011713   & N   & 18.68   & $1681.80\pm437.56$ & $7.55\pm0.56$ \\ \hline
1891     & 5.1640484 & 59.3340478   & V   & 16.62   & $2970.89\pm461.61$ & $4.26\pm0.34$ 
\end{tabular}
\caption{\label{tab:superred2}Details of super red sources detected only in $r$ band. \texttt{V} = non-periodic variable, \texttt{N} = non-variable. Errors on distance and absolute magnitude are calculated from the error in {\it Gaia} parallax measurements.}
\end{table*}

\subsection{Compact Objects}
\label{sec:other foreground}
As shown in Figure \ref{fig:HRdiagram2}, among the ZTF sources with GAIA parallaxes are objects with absolute magnitudes $M_r$ fainter than main sequence objects with the same color (i.e., in  the bottom left corner of the H-R diagram).  Based on this location, many of these are likely to be white dwarfs.

\begin{figure*}
\centering
\includegraphics[width=0.45\linewidth]{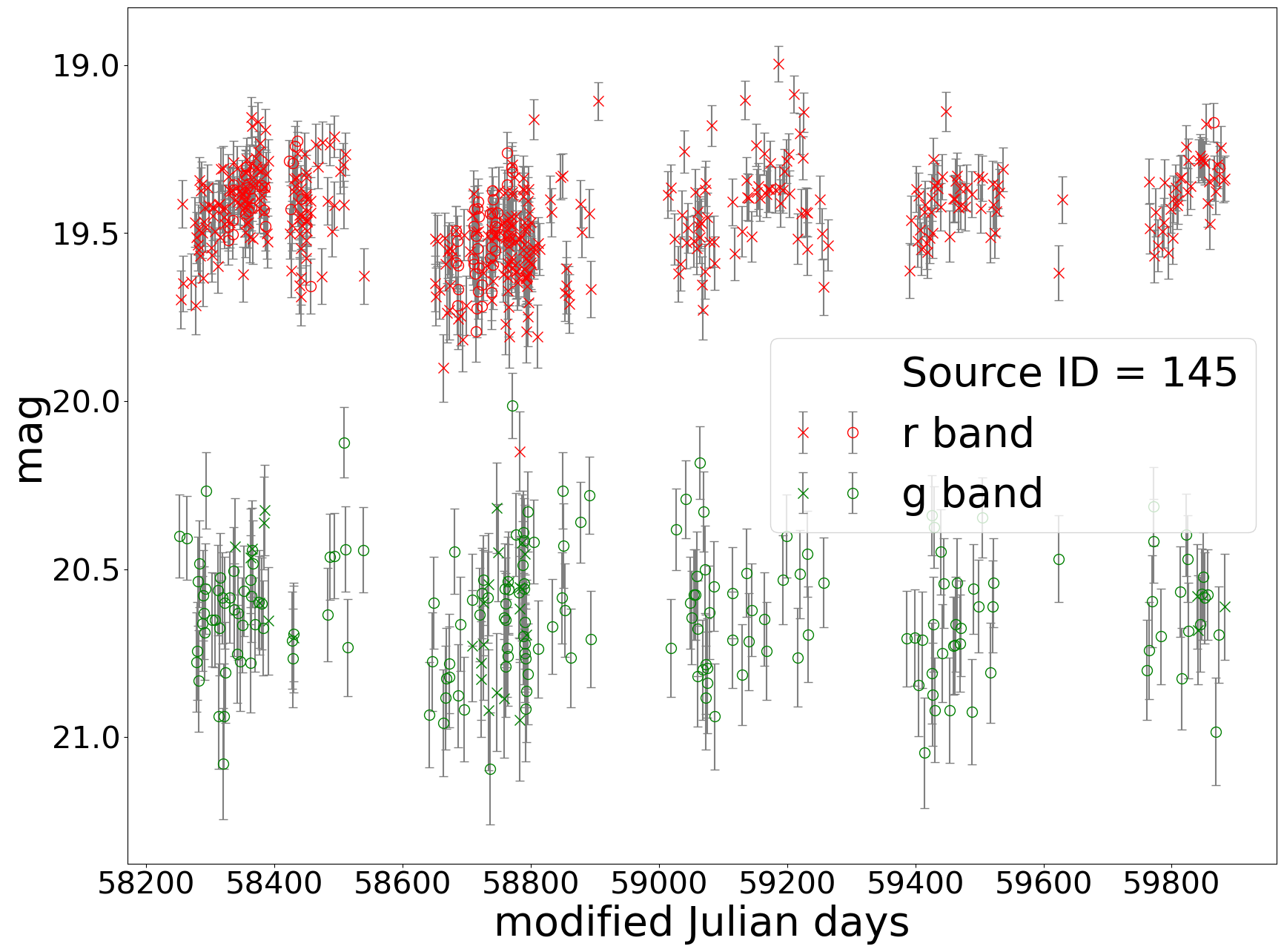}
\includegraphics[width=0.45\linewidth]{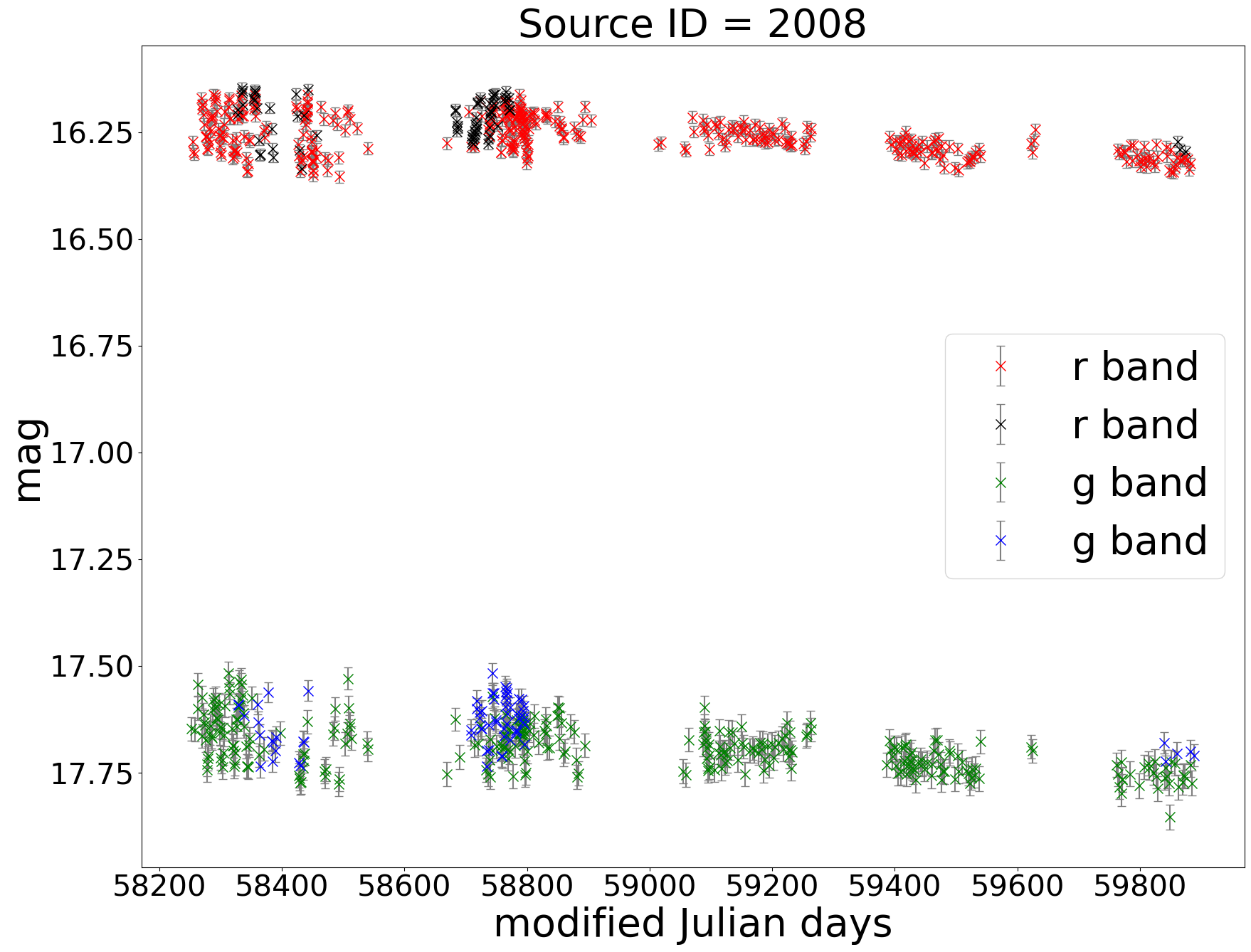}
\includegraphics[width=0.45\linewidth]{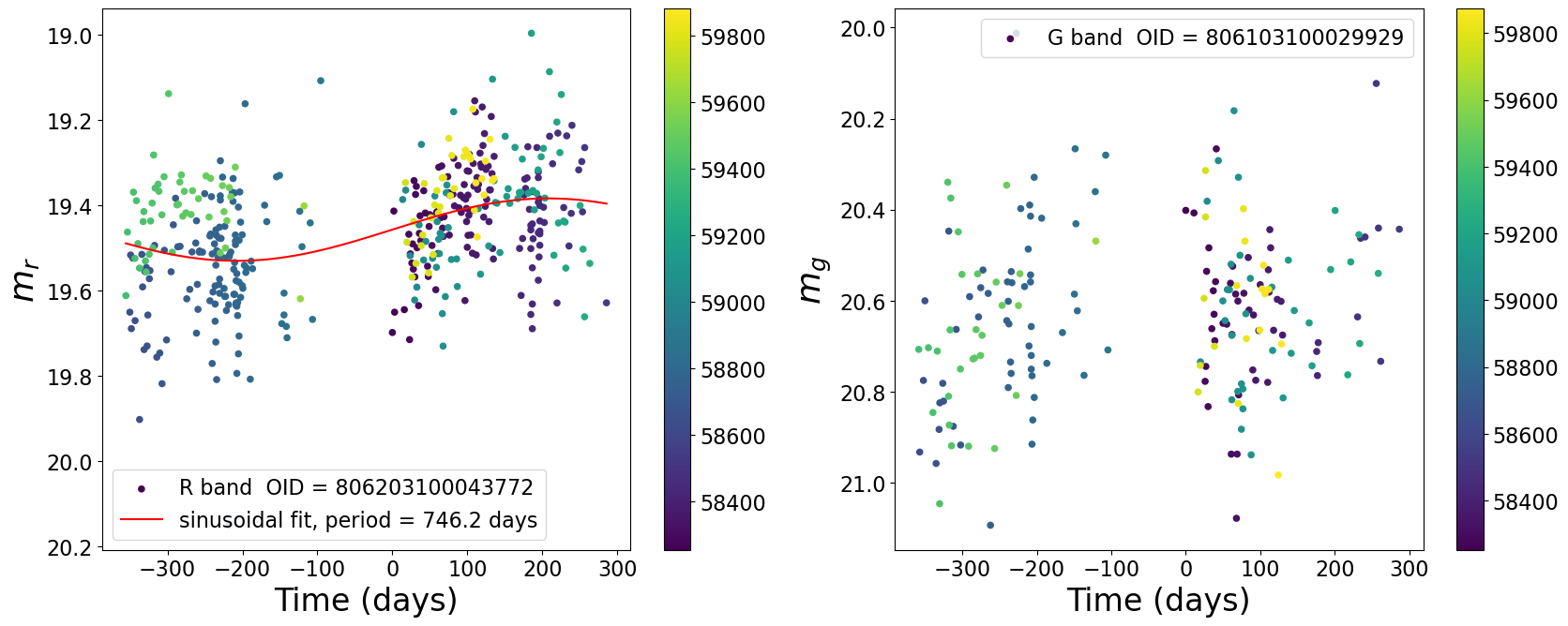}
\hspace{0.5 cm}
\includegraphics[width=0.45\linewidth]{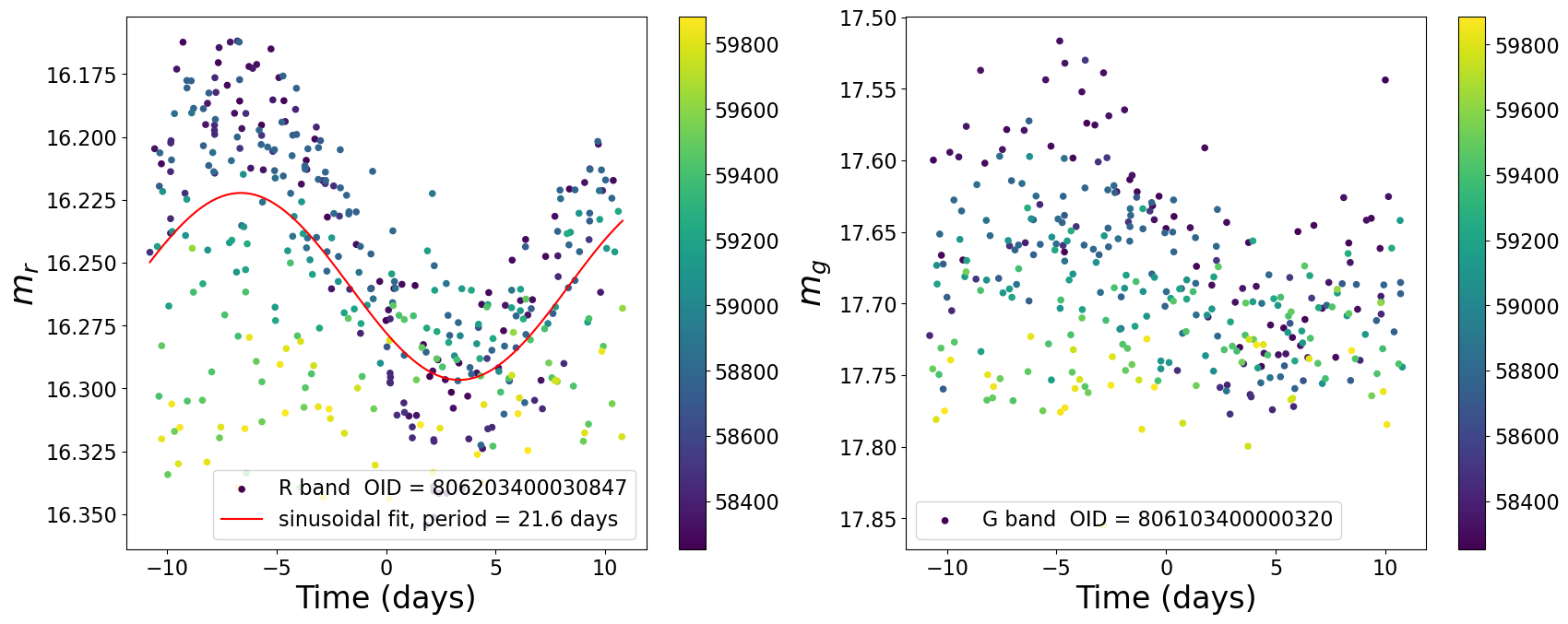}
\caption{Lightcurves of selected compact objects.
\textbf{Left}: Source 145, a possible white dwarf. This source is identified as a periodic variable in $r$ band with a periodicity of 746.2 days and a non-periodic variable in $g$ band. This source is labeled as an emission-line star in SIMBAD.
\textbf{Right}: Source 2008, a possible X-ray binary. This source is identified as a periodic variable in $r$ band with a periodicity of 21.6 days and a non-periodic variable in $g$ band. This source is detected by Chandra.
Different markers represents different ZTF labeled OIDs under the same Source ID.
}
\label{fig:lc_compact_objects}
\end{figure*}

However, the proximity of Source 145 (green triangle in Figure \ref{fig:HRdiagram2}) to the Main Sequence suggests it is a sub-dwarf -- with the SIMBAD identification of this source as an emission line star suggesting it is a hot subdwarf.  Furthermore, as shown in  Figure \ref{fig:lc_compact_objects} ({\it left} panel), its r-band emission periodically varies on a  $\sim750$ days timescale.  Such long periods are rare among hot subdwarfs, and their formation of such binary systems is poorly understood (e.g., \cite{Barlow_2013} and \cite{Deca_2012}).

While not in located in the ``white dwarf'' region of the H-R diagram, we suspect that Source 2008 likely contains a compact object -- either a neutron star or stellar-mass black hole.  The detection of X-ray ({\it Chandra}) emission from this sources and the observed 21.6~day periodicity in its optical emission (right panel of Figure \ref{fig:lc_compact_objects}) suggests that is likely an X-ray binary.   The location of this object in the H-R diagram suggest the stellar companion is like a red giant, making this a probably Galactic low-mass X-ray binary (LMXB).

\section{Conclusion}
\label{sec:conclusion}
In this paper, we used ZTF observations between 2018 Mar and 2022 Nov to identify variable sources in the field of IC 10. We first identified variable lightcurves by measuring the probability that a lightcurve is consistent with a constant magnitude model via survival function $S_k(\chi^2)$ (\S\ref{sec:variabilitycut}).
For variables, we test for periodic changes in brightness using the false alarm probability of the peak power in its Lomb-Scargle periodogram (\S\ref{sec:FAP}), accounting for erroneous periodic behavior due to aliasing with $Q_P$ (\S\ref{sec:Q_P}) and $C_T$ (\S\ref{sec:C_T}). 
We then combine OIDs (ZTF source label) whose positions are $<1^{\prime \prime}$ apart into the same source ID (\S\ref{sec:group}).
We therefore classify all sources in this field into three main categories: non-variable, non-periodic variable, and periodic variable. 
We verified the robustness of the classification by cross-checking our results with a larger ZTF variable catalog by \cite{Chen_2020} (\S\ref{sec:chen}), and compared our methodology with results derived from the machine learning powered classification scheme UPSILoN (\S\ref{sec:upsilon}).

We then studied the demographics of the source classification to better understand the population of sources (\S\ref{sec:demo}). As shown in Figure~\ref{fig:cmd}, variable sources are only identified when $m_r \lesssim 20$, suggesting this is the faintest magnitude for which this work can detect variability.
To identify the physical nature of these sources, we cross-matched the positions of the ZTF sources with SIMBAD -- finding a wide variety of object types identified in existing literature (\S\ref{sec:simbadmatch}).

Based on Gaia parallax and ZTF source density, we distinguished between sources likely physically associated with IC 10 from sources that are not (\S\ref{sec:gaia}). 
For IC 10 sources, we identified several flaring super giants (\S\ref{sec:ic10-flares}), a candidate luminous blue variable (LBV) with a long secondary period (LSP) (\S\ref{sec:193}), six periodic super giants including a possible S Doradus LBV (\S\ref{sec:periodic_sg}), and three possible Mira variables (\S\ref{sec:superred}).  Follow-up observations of these source could help improve distance estimates to this galaxy, as well as better understand the impact of its evolved massive star population on its ISM.

For non-IC 10 sources we identify some super red sources (\S\ref{sec:superred2}), as well as a few compact objects such as a periodic subdwarf and a low mass X-ray binary in the Milky Way (\S\ref{sec:other foreground}).

\acknowledgments
{This material is based upon work supported by Tamkeen under the NYU Abu Dhabi Research Institute grant CASS. Based on observations obtained with the Samuel Oschin 48-inch Telescope at the Palomar Observatory as part of the Zwicky Transient Facility project. ZTF is supported by the National Science Foundation under Grant No. AST-1440341 and a collaboration including Caltech, IPAC, the Weizmann Institute for Science, the Oskar Klein Center at Stockholm University, the University of Maryland, the University of Washington, Deutsches Elektronen-Synchrotron and Humboldt University, Los Alamos National Laboratories, the TANGO Consortium of Taiwan, the University of Wisconsin at Milwaukee, and Lawrence Berkeley National Laboratories. Operations are conducted by COO, IPAC, and UW. 

The data and code used for this work are available for download from the following GitHub repository: \url{https://github.com/ZehaoJin/Transients-in-IC-10}.
}

\bibliography{bibliography.bib}{}
\bibliographystyle{aasjournal.bst}

\end{document}